\documentclass{aa}
\usepackage{graphicx}		% needed for the figures
\usepackage{multicol}
\usepackage[colorlinks = true,
            linkcolor = blue,
            urlcolor  = blue,
            citecolor = blue,
            anchorcolor = blue]{hyperref}
\usepackage{epstopdf}
\usepackage{color}
\usepackage{amsmath}
\usepackage{float}		% this is to place figures where requested!
\usepackage{longtable}
\usepackage{amssymb}
\usepackage{textcomp}
\usepackage{gensymb}
\usepackage[procnames]{listings}
\usepackage{tabto}
\usepackage{paralist} % Used for the compactitem environment which makes bullet points with less space between them
\usepackage{mathtools}
\usepackage{relsize}
\usepackage[normalem]{ulem}
\usepackage{xcolor}
\DeclareGraphicsExtensions{.pdf,.png,.jpeg}

\def\sgra{Sgr\,A$^{\ast}$}
\def\rgauss{$\mathcal{R}_{\boldsymbol \Sigma}$}
\newcommand{\lsum}{\mathlarger{\sum}}

\def\apjl{ApJL}%           % Astrophysical Journal, Letters 
\def\apjs{ApJS}%           % Astrophysical Journal, Supplement 
\def\ao{Appl.~Opt.}%           % Applied Optics 
\def\aap{A\&A}%           % Astronomy and Astrophysics 
\newcommand{\radboud}{Department of Astrophysics/IMAPP, Radboud University, P.O. Box 9010, 6500 GL Nijmegen, The Netherlands}
\newcommand{\cfa}{Center for Astrophysics $|$ Harvard \& Smithsonian, 60 Garden Street, Cambridge, MA 02138, USA}

\begin{document}

%commands for equations
\newcommand{\Ik}{\ensuremath{I_\mathrm{i}}}
\newcommand{\Ij}{\ensuremath{I_\mathrm{j}}}
\newcommand{\xo}{\ensuremath{\bar{x}}}
\newcommand{\xk}{\ensuremath{x_\mathrm{i}}}
\newcommand{\yk}{\ensuremath{y_\mathrm{i}}}
\newcommand{\xj}{\ensuremath{x_\mathrm{j}}}
\newcommand{\yj}{\ensuremath{y_\mathrm{j}}}
\newcommand{\yo}{\ensuremath{\bar{y}}}
\newcommand{\sigxx}{\ensuremath{\Sigma_\mathrm{xx}}}
\newcommand{\sigxy}{\ensuremath{\Sigma_\mathrm{xy}}}
\newcommand{\sigyx}{\ensuremath{\Sigma_\mathrm{yx}}}
\newcommand{\sigyy}{\ensuremath{\Sigma_\mathrm{yy}}}
\newcommand{\sumk}{\ensuremath{\sum\limits_\mathrm{k}}}
\newcommand{\lmaj}{\ensuremath{\lambda_\mathrm{maj}}}
\newcommand{\lmin}{\ensuremath{\lambda_\mathrm{min}}}
\newcommand{\xmaj}{\ensuremath{\theta_\mathrm{maj}}}
\newcommand{\xmin}{\ensuremath{\theta_\mathrm{min}}}

\nocite{PaperI}
\nocite{PaperII}
\nocite{PaperIII}
\nocite{PaperIV}
\nocite{PaperV}
\nocite{PaperVI}

\title{VLBI imaging of black holes via second moment regularization}

\author{S.~Issaoun\inst{1,2} \and M.~D.~Johnson\inst{2}, 
        L.~Blackburn\inst{2} \and M.~Mo{\'s}cibrodzka\inst{1}
      \and A.~Chael\inst{2} \and
      H.~Falcke\inst{1}
      }

\institute{\radboud
     \and
         \cfa 
         }

\abstract{The imaging fidelity of the Event Horizon Telescope (EHT) is currently determined by its sparse baseline coverage. 
In particular, EHT coverage is dominated by long baselines, and is highly sensitive to atmospheric conditions and loss of sites between experiments. The limited short/mid-range baselines especially affect the imaging process, hindering the recovery of more extended features in the image. We present an algorithmic contingency for the absence of well-constrained short baselines in the imaging of compact sources, such as the supermassive black holes observed with the EHT. This technique 
enforces a specific second moment on the reconstructed image in the form of a size constraint, which corresponds to the curvature of the measured visibility function at zero baseline. The method enables the recovery of information lost in gaps of the baseline coverage on short baselines and enables corrections of any systematic amplitude offsets for the stations giving short-baseline measurements present in the observation. 
The regularization can use historical source size measurements to constrain the second moment of the reconstructed image to match the observed size. We additionally show that a characteristic size can be derived from available short-baseline measurements, extrapolated from other wavelengths, or estimated without complementary size constraints  with parameter searches. 
We demonstrate the capabilities of this method for both static and movie reconstructions of variable sources. }


\maketitle 

\section{Introduction}\label{sec:intro}

Very-long-baseline interferometry (VLBI) is a technique able to achieve high angular resolution imaging through the use of widely separated antennas. Unfortunately, as the observing frequency is increased, the availability of suitable sites on Earth is greatly reduced, leading to sparse arrays with a high angular resolution but a low spatial dynamic range. 
In particular, a simple inverse Fourier transform of the visibilities measured by an interferometer, or `dirty image', is dominated by artifacts introduced by sparse sampling of the Fourier plane. Short baselines are particularly important in imaging, as they anchor the flux distribution and provide a crucial link between high-resolution small-scale features and the large-scale extent and morphology of the target. The sparser the array, the more challenging it is to reconstruct images from interferometric measurements. Additionally, weather and technical issues at sites that provide short/mid-range baselines can greatly
degrade the ability to image a given data set. 

Array sparsity and station-based errors can have dramatic effects on reconstructed images. Thus, the imaging process requires further information and assumptions beyond the visibility measurements from the interferometer. The choice of imaging method imposes additional constraints on the reconstructed image. Here, we will focus on extending the method of regularized maximum likelihood (RML) that performs well under sparse sampling conditions and does not involve direct inverse Fourier transforms of the data in the imaging process. 

In this paper we present an algorithmic contingency to array sparsity and site issues in the form of a second moment regularization function. That is, the compactness of the source can be expressed as the second moment of the source brightness distribution \citep{moffet_1962,burn_1976}, which can be constrained to match, for example, confident source size measurements from short baselines of previous experiments or epochs. Enforcing this source size constraint supplements limited short-baseline information while fitting to long-baseline smaller scale structure from newer observations. 

The Event Horizon Telescope (EHT), observing at a frequency of 230\,GHz \citep{PaperI,PaperII}, is a prime example of a high-frequency VLBI imaging experiment with image uncertainties dominated by the effects of sparse coverage. The EHT currently has only a single short/mid-range VLBI baseline, joining the Large Millimeter Telescope Alfonso Serrano (LMT) in Mexico to the Submillimeter Telescope (SMT) in Arizona. Recent observations with the EHT have shown that the LMT is difficult to calibrate, giving baselines with large measurement uncertainties dominated by uncharacterized station behavior in 2017~\citep{PaperIII,PaperIV}.

Although the EHT observes a number of non-horizon-scale sources in conjunction with the Atacama Large Millimeter/submillimeter Array (ALMA), its primary targets are the two supermassive black hole candidates in the Galactic Center, Sagittarius A* (\sgra), and at the center of the radio galaxy M87. At the frequency of the EHT, these two sources are very compact, with sizes on the sky historically measured with three stations, in California, Arizona, and Hawaii, in early EHT observations, and are thus ideal imaging targets for second moment regularization~\citep{Doeleman_2008,Fish_2011,Doeleman_2012,Akiyama_2015,Johnson_2015,Lu_2018}. Near-zero closure phases on the California--Arizona--Hawaii triangle are indicative of the source compactness and symmetry on scales of a few tens of $\mu$as \citep{Akiyama_2015,Fish_2016}. The California--Arizona baseline provided the short-baseline measurements needed to constrain the compactness and size of the sources in the visibility domain. Recent observations of M87 in 2017 also found a source size of $\sim40\,\mu$as consistent with previous measurements \citep{PaperI,PaperII,PaperIII,PaperIV,PaperV,PaperVI}. 

For \sgra, the source size is also well-constrained at lower frequencies due to its compactness and dominant diffractive scattering \citep{Shen_2005,Bower_2006,Lu_2011,Johnson_2018}. VLBI observations at 86\,GHz taken one month apart give fitted Gaussian source sizes for the scattered image of \sgra with $<10\%$ difference \citep{Ortiz_2016,Brinkerink_2019}. At this frequency, while the small scale structure is expected to vary, the large-scale information, dominated by the size of the scattering kernel, should be stable from epoch to epoch~\citep{Johnson_2018}. 

Second moment regularization merges the benefits of model-fitting with the flexibility of imaging: compared to self-calibration to a known model, it does not actually modify the measured visibilities used for the imaging process or enforce a model-dependent solution, but instead provides additional information to improve image quality. The regularization constrains the spread of flux density to a motivated region in the image, discouraging non-physical morphology driven by fits to long-baseline data and accelerating convergence toward a plausible image. It is a natural extension of imaging tools that add source information in the imaging process in RML methods: a total flux constraint is in fact the zeroth moment of the image; an image centroid specification corresponds to the first moment of the image; and a short-baseline source size completes the picture by constraining the image second moment. The implementation of second moment regularization can be done in conjunction with other tools and constraints in RML, for both static and movie reconstructions. 
Furthermore, as the constraint function acts on the image itself and does not modify the visibility data, it can be used with any choice of data product, including minimally-calibrated closure phases and amplitudes. 

The paper is structured as follows. We present the mathematical background to motivate the regularization in Sect.~\ref{sec:background}. We outline the method, assumptions, and physical motivation in Sect.~\ref{sec:method}. In Sect.~\ref{sec:demo} we demonstrate the improvements in image quality and fidelity using the regularization with or without a priori knowledge of the source size. We present possible applications of the second moment regularization to more sophisticated imaging techniques for scattering mitigation and movie reconstructions in Sect.~\ref{sec:examples}. A summary is given in Sect.~\ref{sec:summary}.

\section{Background}\label{sec:background}
By the van Cittert-Zernike theorem, an interferometer samples complex visibilities corresponding to Fourier components of an image \citep{vancittert,zernike}. Consequently, $n^{\text{th}}$ moments of an image correspond to $n^{\text{th}}$ derivatives of the visibility function at the origin. 
Specifically, an interferometric visibility $V(\mathbf{u})$ on a baseline $\mathbf{u}$ can be written as \citep[e.g.,][]{TMS}
\begin{align}
V(\mathbf{u}) &= \int d^2\mathbf{x}\, I(\mathbf{x}) e^{-2\pi i \mathbf{u} \cdot \mathbf{x}}, \label{eq:cittert-zernicke}
\end{align}
where $I(\mathbf{x})$ is the brightness distribution on the sky, and $\mathbf{x}$ is an angular unit. 

From this expression, $V(\mathbf{0}) = \int d^2\mathbf{x}\, I(\mathbf{x}) \in \mathbb{R}$ gives the total flux density of the image (the $0^{\text{th}}$ moment). Likewise, the phase gradient of the visibility function at zero baseline gives a vector proportional to the centroid of the image,
\begin{align}
\nonumber \left. \nabla V(\mathbf{u}) \right\rfloor_{\mathbf{u}=\mathbf{0}} &= -2\pi i \int d^2\mathbf{x}\, \mathbf{x} I(\mathbf{x}) \\
&= -2\pi i V(\mathbf{0}) \boldsymbol{\mu},
\end{align}
where $\boldsymbol{\mu}$ is the image centroid (the normalized $1^{\text{st}}$ moment):
\begin{align}
\boldsymbol{\mu} = (\xo \mathbf{\hat{x}}, \yo \mathbf{\hat{y}}) = \frac{\int d^2\mathbf{x} I(\mathbf{x}) \mathbf{x}}{\int  d^2\mathbf{x}\, I(\mathbf{x}) }.
\end{align}
Because the image is real, the gradient $\left. \nabla V(\mathbf{u}) \right\rfloor_{\mathbf{u}=\mathbf{0}}$ is purely imaginary. For images that are positive (e.g., images in total intensity), the visibility function must take its maximum amplitude at the origin. More generally, the visibility function is Hermitian; thus, its amplitude must always have a vanishing gradient at the origin because of the conjugation symmetry $V(\mathbf{u}) = V^\ast(-\mathbf{u})$. 

The second derivative, or Hessian, of the visibility amplitude function at zero baseline gives a matrix (see Appendix~\ref{sec:vis_der}):
\begin{align}
\nonumber \left. \nabla\nabla ^\intercal |V(\mathbf{u})| \right\rfloor_{\mathbf{u}=\mathbf{0}} &= -4\pi^2 \int d^2\mathbf{x}\, I(\mathbf{x})(\mathbf{x}-\boldsymbol{\mu}) (\mathbf{x}-\boldsymbol{\mu})^\intercal \\
    &= -4\pi^2 V(\mathbf{0}) \boldsymbol{\Sigma}, \label{eq:2mom}
\end{align}
where $\boldsymbol{\Sigma}$ is the normalized second central moment (or covariance matrix) of the image. 
We show in Appendix~\ref{sec:vis_der} that this expression is equivalent to the curvature of the centered complex visibility function \citep[see also][]{moffet_1962,burn_1976}. 
The visibility amplitude function is a more natural data product to use for observations with non-astrometric VLBI arrays such as the EHT, where there is no absolute phase information due to strong differential atmospheric propagation effects between sources, and thus no directly measured full complex visibilities. Therefore it is useful for us to determine image moments directly from the visibility amplitude function, which is measured. 

\begin{table*}[t]
    \centering
    \setlength{\tabcolsep}{8pt} 
    \caption{Correspondence of the mass, center of mass and moment of inertia in the image and visibility domains.}
    \begin{tabular}{lllll}
    \hline 
    Physical Analog & \multicolumn{2}{l}{Image Domain} & \multicolumn{2}{l}{Visibility Domain}\\
    \hline 
    \hline 
    Mass & Total Flux & $\int I(\mathbf{x}) d^2\mathbf{x}$ & Peak Visibility & $V(\mathbf{0})$ \\
    Center of Mass & Centroid ($\boldsymbol{\mu}$) &  $V(\mathbf{0})^{-1} \int \mathbf{x} I(\mathbf{x}) d^2\mathbf{x}$ & Phase Gradient &  $(2\pi i V(\mathbf{0}))^{-1} \left. \nabla V(\mathbf{u}) \right\rfloor_{\mathbf{u}=\mathbf{0}}$ \\
    Moment of Inertia & Covariance ($\boldsymbol{\Sigma}$) & $V(\mathbf{0})^{-1} \int \mathbf{x} \mathbf{x}^\intercal I(\mathbf{x}) d^2\mathbf{x}$ & Amplitude Curvature & $(-4\pi^2 V(\mathbf{0}))^{-1} \left. \nabla\nabla ^\intercal V(\mathbf{u}) \right\rfloor_{\mathbf{u}=\mathbf{0}} $ \\
    \hline 
    \end{tabular}
    \label{tab:inertia}
\end{table*}

The image covariance matrix $\boldsymbol{\Sigma}$ can be more intuitively expressed in terms of its principal axes, corresponding to the perpendicular axes about which the second moment reaches its maximum \citep{Hu_1962}. The matrix has two eigenvalues $\lmin$ and $\lmaj$, and can be diagonalized as follows:
\begin{align}
\mathbf{\Sigma} 
= \mathbf{R_\phi} \begin{pmatrix} 
\lmin & 0 \\ 
0 & \lmaj
\end{pmatrix} \mathbf{R_\phi^{\intercal}},  
\end{align}
where $\mathbf{R_\phi}$ is the rotation matrix based on the position angle east of north $\phi$  of the major principal axis (Appendix \ref{sec:vis_axes}).
The eigenvalues of the covariance matrix are the variances of the normalized image projected along the principal (major and minor) axes. The correspondence between \lmaj, \lmin, $\phi$ and the individual terms of $\boldsymbol{\Sigma}$ is given in Appendix \ref{sec:vis_axes}. 

Following Equation~\ref{eq:cittert-zernicke}, we can fully express the visibility function as a Taylor expansion in its derivatives.
Each $n+1^\text{th}$ term of the Taylor expansion is proportional to the $n^\text{th}$ moment of the visibility function (see Table~\ref{tab:inertia}). At zero baseline, only the zeroth moment remains. We choose the coordinate system such that the centroid of the image is at the origin, and the first moment of the visibility function (the second term of the Taylor expansion) vanishes. At short baseline, the centered complex visibility function is therefore dominated by the quadratic term. The Taylor expansion of the visibility function at short baseline becomes:
\begin{align}
\nonumber V(\mathbf{u}) &\simeq V(\mathbf{0}) - 2\pi^2  \int d^2\mathbf{x}\, (\mathbf{u}\cdot \mathbf{x})^2 I(\mathbf{x}) \\
    &\simeq V(\mathbf{0}) - 2\pi^2 V(\mathbf{0})\mathbf{u}^\intercal \boldsymbol{\Sigma} \mathbf{u} . \label{eq:gen-short}
\end{align}
 
 Equation~\ref{eq:gen-short} describes the visibility function behavior on short baselines entirely in terms of the total flux $V(\mathbf{0})$ and the second moment covariance matrix $\boldsymbol{\Sigma}$ projected along the baseline direction. These parameters also describe a unique visibility function of a Gaussian source with total flux $V(\mathbf{0})$, and major/minor axes sizes and orientation prescribed by the same covariance matrix. We show this by comparing the general complex visibility function to that for a Gaussian source. For the simplest case of an isotropic Gaussian source of standard deviation $\sigma$ with the same total flux $V(\mathbf{0})$, we have the following intensity pattern on the sky and corresponding visibility function:
\begin{align}
 I_\mathrm{gauss}(\mathbf{x}) &=  \frac{V(\mathbf{0})}{2\pi\sigma^2} e^{-|\mathbf{x}|^2/2\sigma^2}, \\
 V_\mathrm{gauss}(\mathbf{u}) &=  V(\mathbf{0})e^{-2\pi^2|\mathbf{u}|^2\sigma^2} .
\end{align}
More generally, an anisotropic Gaussian with a covariance matrix $\boldsymbol{\Sigma}$ gives:
\begin{align}
 I_\mathrm{gauss}(\mathbf{x}) &=  \frac{V(\mathbf{0})}{2\pi\sqrt{|\boldsymbol{\Sigma}|}} e^{-\mathbf{x}^\intercal \boldsymbol{\Sigma}^{-1}\mathbf{x}}, \\
V_\mathrm{gauss}(\mathbf{u})  &=  V(\mathbf{0})e^{-2\pi^2\mathbf{u}^\intercal \boldsymbol{\Sigma}\mathbf{u}}.
\end{align}
Taking the Taylor expansion of the anisotropic Gaussian visibility function at short baselines, the first two terms dominate:
\begin{align}
V_\mathrm{gauss}(\mathbf{u}) \simeq V(\mathbf{0}) - 2\pi^2V(\mathbf{0}) \mathbf{u}^\intercal \boldsymbol{\Sigma}\mathbf{u} .\label{eq:gauss-short}
\end{align}
We thus obtain an equivalence of the behavior of the general visibility function (Equation~\ref{eq:gen-short}) and the Gaussian visibility function (Equation~\ref{eq:gauss-short}) at short baselines. This relation allows us to translate the second moment covariance matrix of the general visibility function to the covariance matrix of an anisotropic Gaussian, which provides a simple parametrization to describe the second moment in terms of the characteristic source extent. The sizes of the major and minor axes \xmaj\ and \xmin\ are simply the full widths at half-maximum (FWHMs) of the equivalent Gaussian derived from the variances projected along each principal axis: 
\begin{align}
\xmaj &= \sqrt{8\ln(2)\lmaj}, \\
\xmin &= \sqrt{8\ln(2)\lmin}.
\end{align}

\begin{figure}[t]
\includegraphics[width=\linewidth]{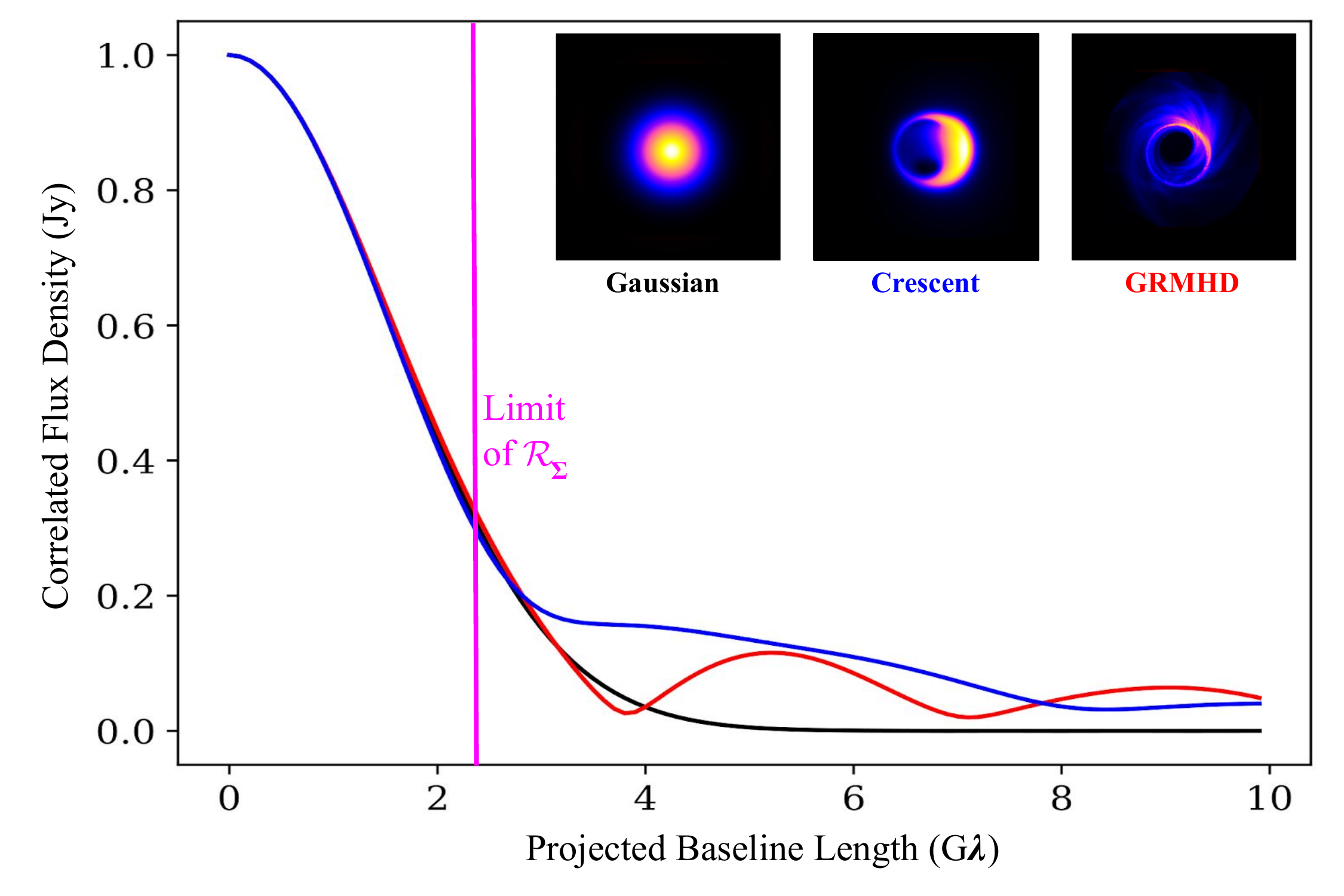}
\caption{Three images with equal extent along their respective major axis, 
 from left to right: a Gaussian; a crescent model; a ray-traced image from a general relativistic magneto-hydrodynamics (GRMHD) simulation of a black hole shadow and accretion disk. Model visibility amplitudes along the major axis of each source as a function of $(u,v)$ distance, after flux and size normalization, show identical behavior at short baseline length but diverge at longer baseline length: the Gaussian in black; the crescent in blue; and the GRMHD simulation in red. }
\label{fig:behavior}
\end{figure}

The equivalence to the Gaussian also gives a natural break-off point where the characteristic source size constraint from the second moment ceases to be a good approximation to the full visibility function: the $1/e$ point determining the resolvability of a Gaussian translates to the baseline length at which the visibility amplitude reaches $V(\mathbf{0})/e$.  
Baseline lengths longer than the $1/e$ point will lead to higher order terms of the Taylor expansion dominating the behavior and sampling finer structure in the image.
We employ the $1/e$ point as a conceptual and visual limit for the source size constraint applied via the second moment regularization. It is not a hard cut-off enforced by the imaging process. 

In Fig.~\ref{fig:behavior}, we demonstrate the behavior of the normalized visibility amplitudes sampled along the source major axis as a function of projected baseline length for three images with distinctly different structure but an identical second moment. The behavior on short baselines aligns well for all three images, the amplitudes start to diverge at longer baselines. We denote the $1/e$ limit, corresponding to the resolvability of the Gaussian image, with a magenta vertical line. On baselines past this line, the amplitudes show very different behavior, dominated by the smaller-scale features in each image (or lack thereof). We can thus express the visibility amplitude function behavior on short baselines via the second moment of the image, defined by the total flux and just three Gaussian parameters: the principal axes FWHMs \xmaj\ and \xmin\ and the position angle $\phi$ of the major axis east of north. In the RML imaging process, there is an additional fifth input parameter, governing the weight of the second moment regularization, or hyperparameter $\beta_{\textrm{R}}$, following Equation~\ref{eq::objfunc}. 

\section{Method}\label{sec:method}
RML focuses on pixelized reconstructions of the image, iteratively maximizing an ``objective function'', which is analogous to a log posterior probability function. This function is a weighted (via ``hyperparameters'') sum of both $\chi^2_\textrm{D}$ goodness-of-fit data terms, and regularization functions $S_{\textrm{R}}$, or ``regularizers'', governing specific image properties. In this paper, we use the RML method implemented in the \texttt{eht-imaging} Python library \citep{Chael_2016,Chael_2018}, where the objective function $J(I)$ is minimized via gradient descent, and can be written as: 
\begin{align}
 \label{eq::objfunc}
 J(I) = \lsum_{\mathclap{\text{data terms}}} \alpha_{\textrm{D}} \chi^2_{\textrm{D}}\left(I\right) - \lsum_{\mathclap{\text{regularizers}}} \beta_{\textrm{R}} S_{\textrm{R}}\left(I\right),
\end{align}
where $\alpha_{\textrm{D}}$ and $\beta_{\textrm{R}}$ are the input hyperparameters.

Using only five input parameters to the regularization ($V(\mathbf{0})$, \xmaj, \xmin, $\phi$ and $\beta_{\textrm{R}}$) we can now constrain the second moment of the reconstructed image to match the size constraint provided by the user for RML imaging. In Sect.~\ref{sec:2mom} we present our implementation of the second moment regularization function within the \texttt{eht-imaging} library minimization framework. In Sect.~\ref{sec:assumptions} we describe the assumptions and physical motivation for second moment regularization using historical observational measurements, known source properties and theoretical expectations. 

\subsection{Second moment regularization}\label{sec:2mom}

Regularization functions in imaging enforce constraints on particular properties of the image, such as image entropy \citep[e.g.,][]{Narayan_Nityananda_1986}, smoothness \citep{Bouman_2016,Chael_2016, Kuramochi_2018} and/or sparsity \citep{Wiaux_2009a,Wiaux_2009b,Honma_2014,Akiyama_2017b,Akiyama_2017a}. Simple constraints, such as image positivity, image total flux (zeroth moment) or image centering (first moment), are often applied to the image, utilizing known information on the behavior of the total intensity distribution of the source imaged. The implementation of a second moment regularization, constraining the size of the source, is thus a natural extension of common imaging tools that add source information to the imaging process. 

We define a regularization function that is minimized when the covariance matrix of the reconstructed image $\boldsymbol{\Sigma}$ matches a user-specified covariance matrix $\boldsymbol{\Sigma}'$. In practice, this latter matrix is computed using user-specified principal axes FWHMs and position angle. We utilize the Frobenius norm to determine a penalty function that quantifies the difference between the user-specified and reconstructed covariance matrices:
\begin{align}
\mathcal{R}_{\boldsymbol \Sigma} &\equiv  \mathrm{Tr}\left[\left({\boldsymbol \Sigma} - {\boldsymbol \Sigma'}\right)^\intercal \left({\boldsymbol \Sigma} - {\boldsymbol \Sigma'}\right) \right]
\end{align}
This regularizer is, by definition, simply the minimization of the difference between two covariance matrices. The procedure for the regularizer implementation in the {\tt eht-imaging} library via gradient descent is presented in Appendix~\ref{sec:implementation}.

\subsection{Assumptions} \label{sec:assumptions}
The second moment regularization operates under a few key assumptions on the properties of the source observed. The main assumption of this method is the compactness of the source. In order to get a quadratic fall-off in the visibility function, as shown in Sect.~\ref{sec:background}, the source must be compact and resolved on longer baselines of the interferometer. This method would break down for point sources or sources with complex morphology and diffuse flux on large scales. 

Another assumption concerns the stability of the source size across multiple epochs. The input axis sizes and position angle for the regularization will only be valid if the source does not vary significantly in size between observations. The source size input is typically derived from observations where weather conditions, coverage, and station performance on short baselines were adequate for higher precision model fitting. The source size can then be used for data sets with larger uncertainties to improve the fidelity and convergence of the imaging process. 
This assumption is well-motivated for the compact sources observed with the EHT:
\begin{itemize}
    \item \sgra\ at 86\,GHz, has been model-fitted with varying precision over two decades, with little variation in the obtained source size parameters, \citep{Rogers_1994,Krichbaum_1998,Doeleman_2001,Shen_2005,Lu_2011,Ortiz_2016,Brinkerink_2019}
    \item \sgra\ at 230\,GHz has been measured to be compact and stable in size between 2007 and 2013 \citep{Doeleman_2008,Lu_2018, Johnson_2018},
    \item M87 at 230\,GHz has been measured to be compact and stable in size over a decade \citep{Doeleman_2012,Akiyama_2015,PaperI,PaperII,PaperIII,PaperIV,PaperV,PaperVI}.
\end{itemize}
It is worth noting that this assumption breaks down for sources with multiple bright components moving relative to each other, as is common for multi-epoch images of bright jet sources. An overall size measurement from a single epoch would not translate to other observations due to components appearing or moving outward, changing the source morphology significantly between observations. The quadratic fall-off approximation until the $1/e$ point would also not hold for two separated point sources, which do show a quadratic fall-off in the visibility amplitudes but the amplitudes would quickly evolve to more complex structure on longer baselines that could be identified as the behavior of two point sources interfering. The method is most effective whenever the emission is confined within a single compact region or on multiple scales that are substantially separated, and particularly if the scale of the emission in the image is comparable to the resolution of the array. 
 
\begin{figure}[t]
\centering 
\hspace{-0.04\linewidth}\includegraphics[width=0.515\linewidth]{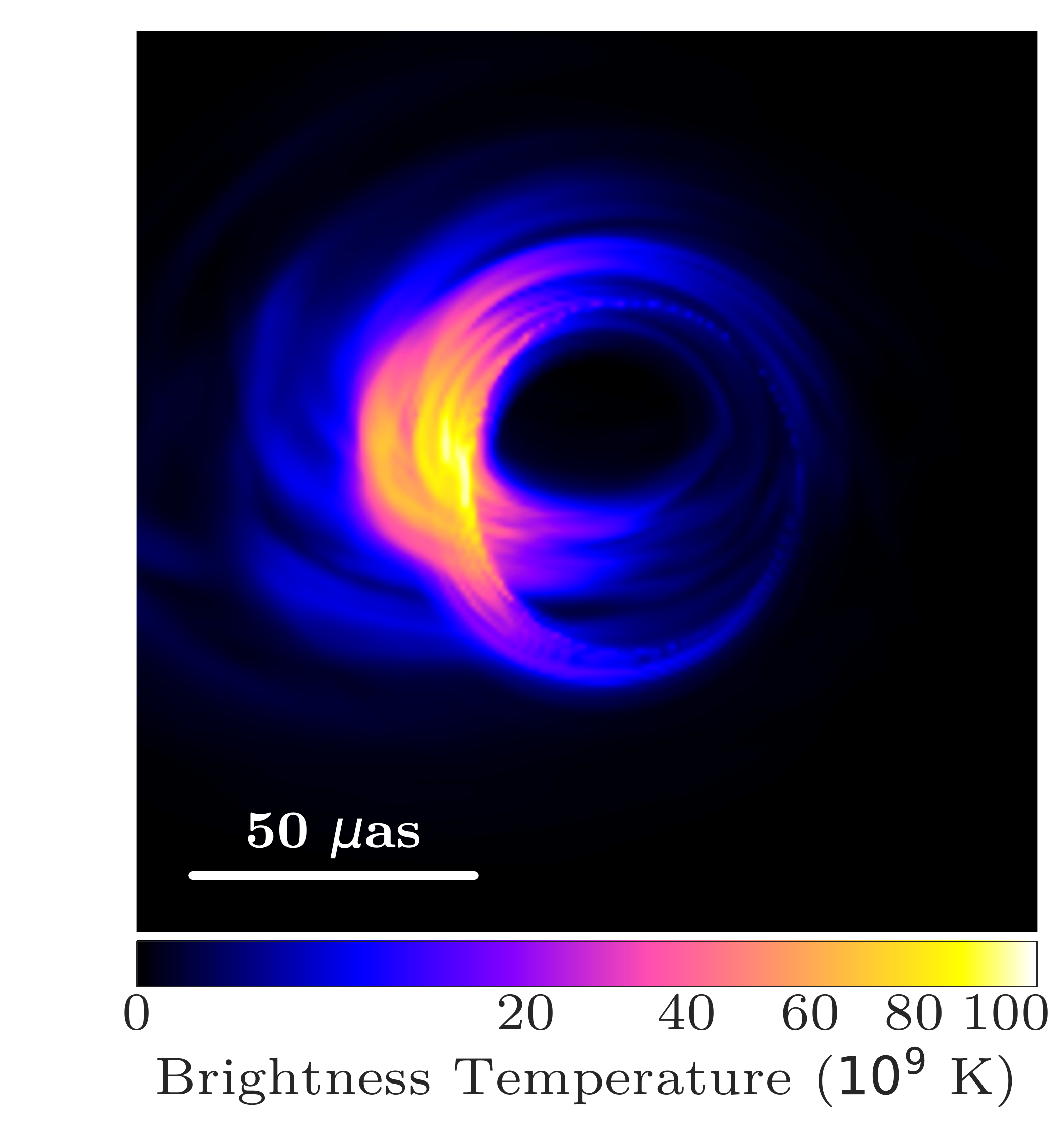}
\includegraphics[width=0.51\linewidth]{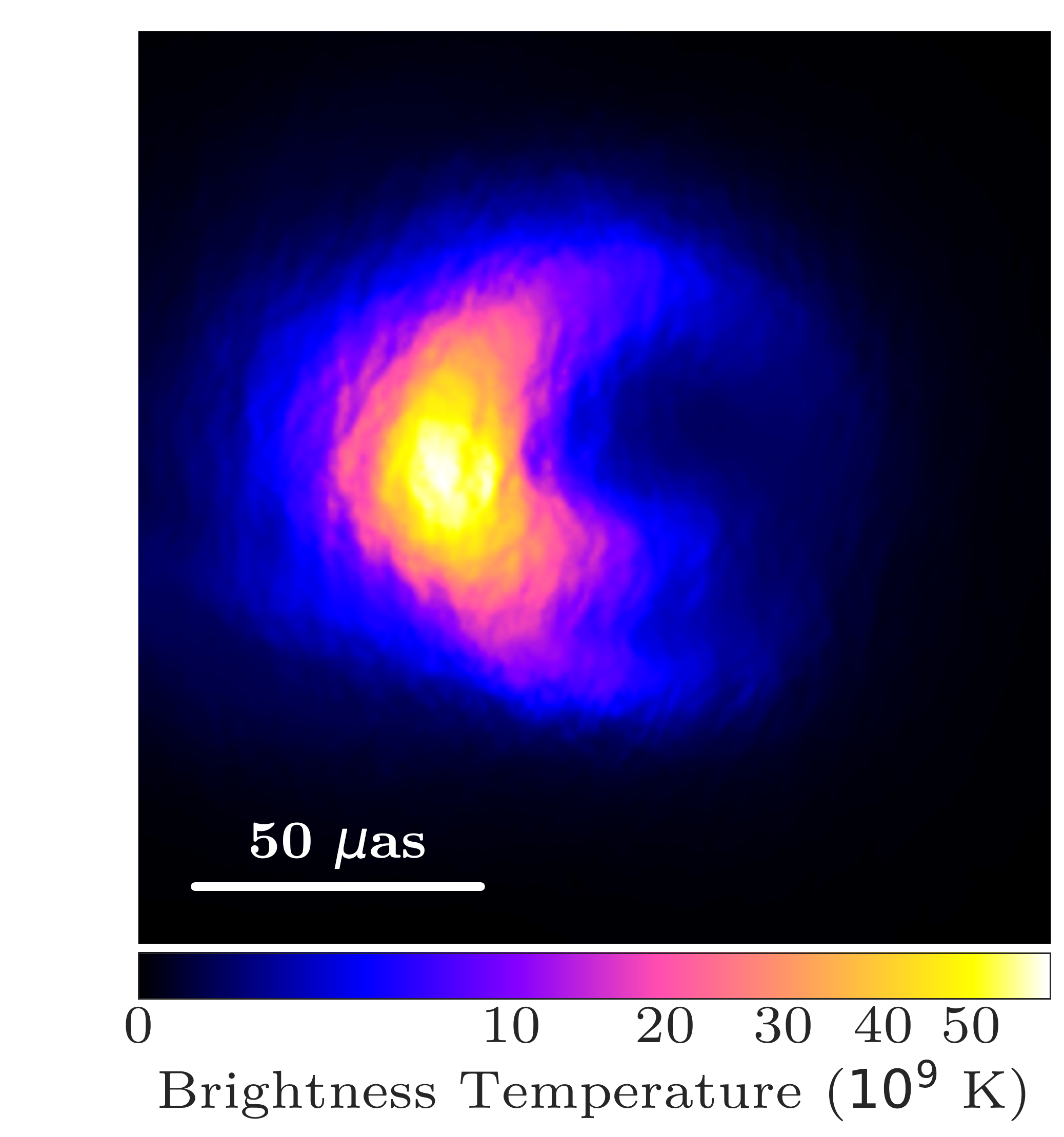}
\caption{Left: 230\,GHz GRMHD simulation of \sgra\ \citep{Moscibrodzka_2018}. Right: Same simulation including the effects of interstellar scattering \citep{Johnson_2016,Johnson_2018}.
}
\label{fig:grmhd_im}
\includegraphics[width=0.95\linewidth]{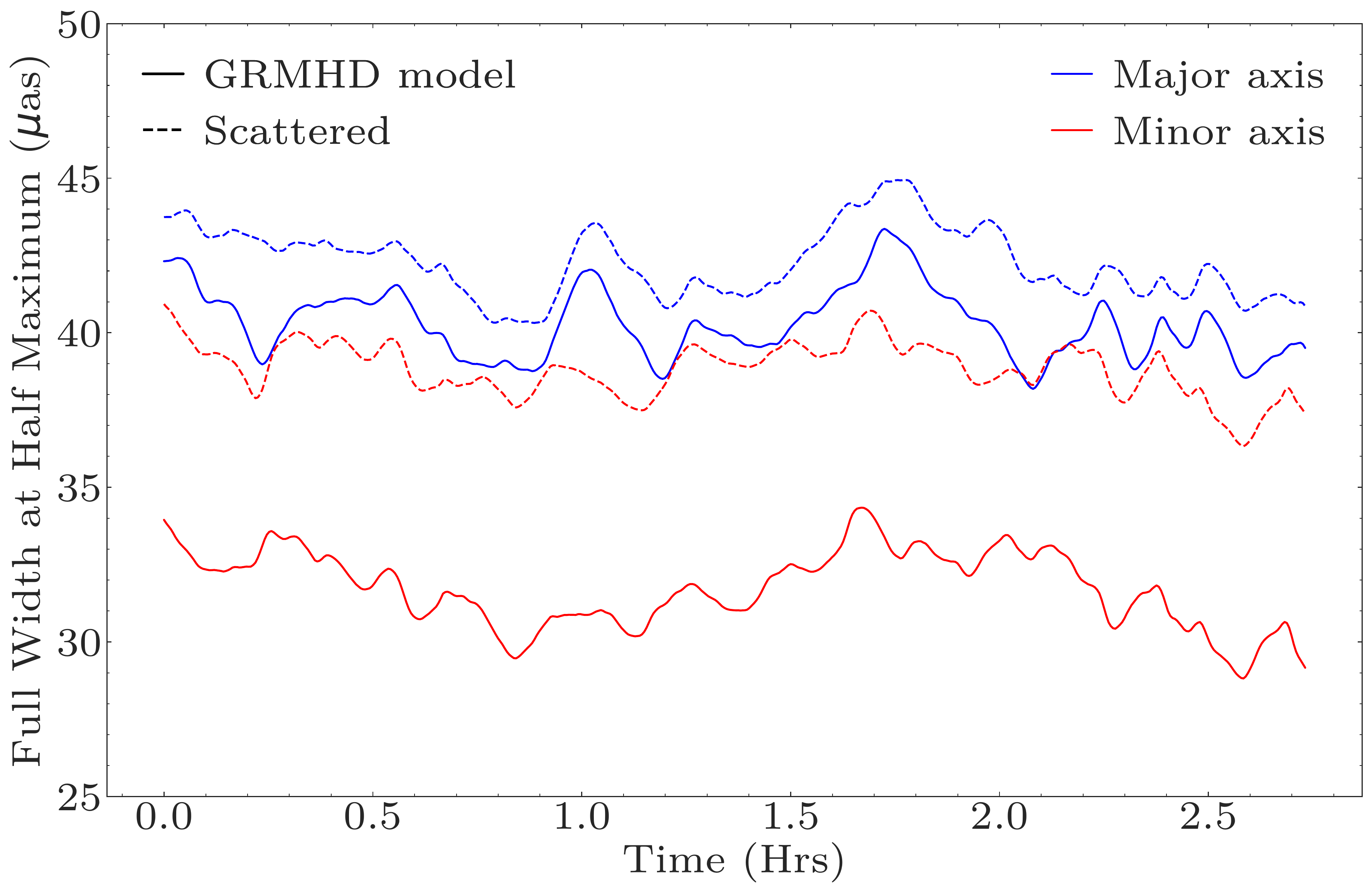}
\caption{Principal axes FWHMs as a function of time for the simulation of \sgra\ in Fig.~\ref{fig:grmhd_im} \citep{Moscibrodzka_2018}. The solid lines show sizes for the simulation, the dotted lines show sizes for the simulation including the effects of interstellar scattering \citep{Johnson_2016,Johnson_2018}. The scattering major axis is aligned with the source minor axis, and thus the scattering kernel slightly dominates the minor axis size, which stabilizes the minor axis FWHM time series. The sizes were obtained from measurements of the image second moment per frame. For all four size trends, the deviation about the mean size is $<10\%$.}
\label{fig:grmhd_var}
\end{figure}

\begin{figure}[t]
\includegraphics[width=0.95\linewidth]{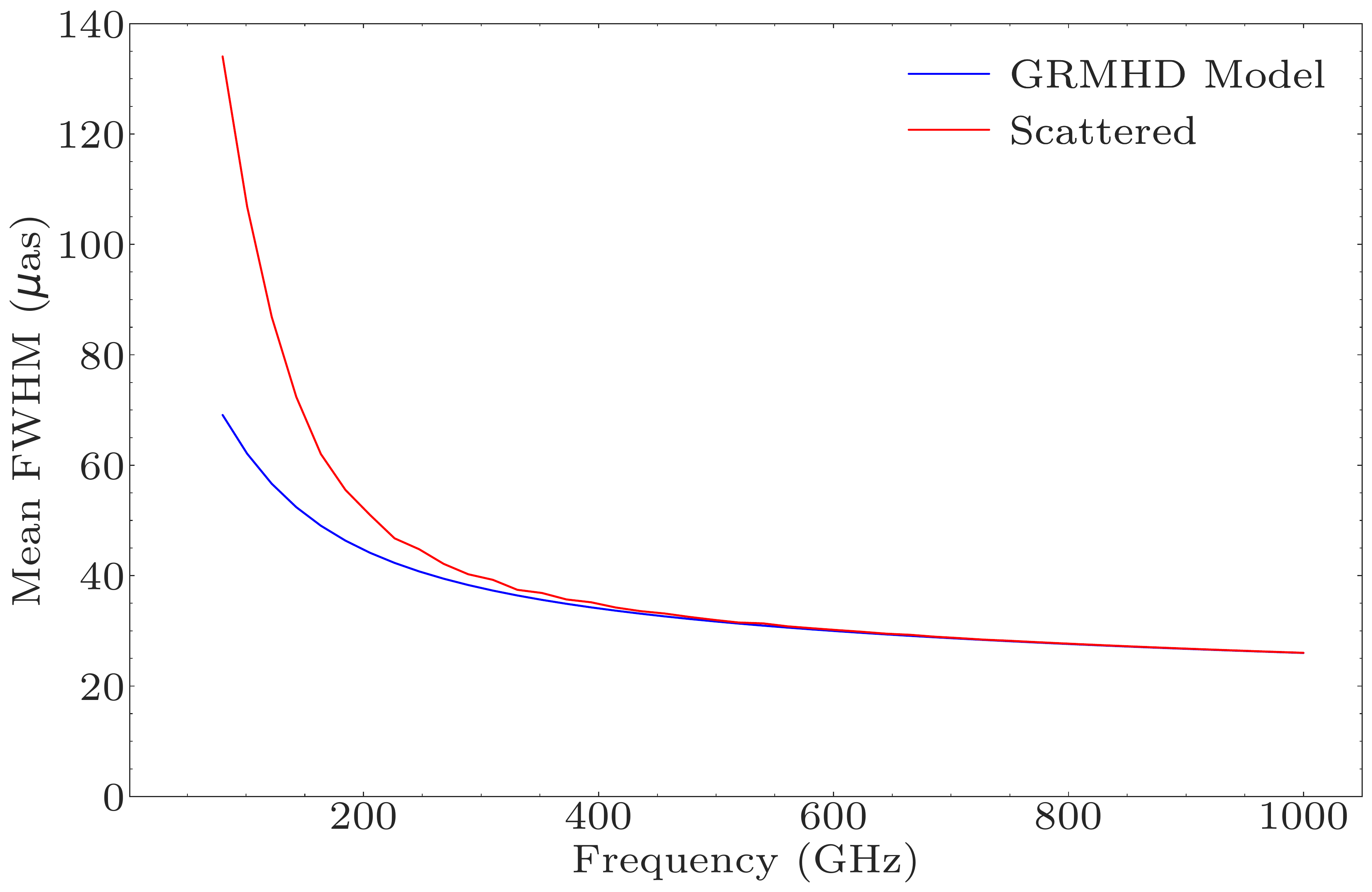}
\caption{Geometric mean FWHM of principal axes as a function of frequency for the ray-traced simulation of \sgra\ in Fig.~\ref{fig:grmhd_im} \citep{Moscibrodzka_2018}. The blue curve shows size evolution for the simulation, the red curve shows size evolution for the simulation including the effects of interstellar scattering \citep{Johnson_2016,Johnson_2018}. The sizes were obtained from measurements of the image second moment per frequency bin of 20\,GHz. The change in size with increasing frequency becomes greatly reduced at frequencies above 300\,GHz, where the size of the source is dominated by the achromatic black hole shadow and the Doppler boosted features \citep{Falcke_2000}.}
\label{fig:grmhd_freq}
\end{figure}

We also assume that the extent of the source does not significantly vary within a single epoch. For static imaging of slow-varying sources, it suffices to assume that the average size of the source matches the input, but this has further implications on reconstructions of variable sources within a single epoch. The structural variability on short timescales should be contained within the region constrained by the second moment. This is an issue particularly for imaging \sgra, as the source is known to vary on timescales of minutes, much shorter than the length of a single observing epoch. We assess the degree of variability of the source extent in quiescent (non-flaring) models of \sgra\ using general relativistic magnetohydrodynamic (GRMHD) simulations of variable emission on horizon scales \citep[Fig.~\ref{fig:grmhd_im};][]{Moscibrodzka_2018}. In Fig.~\ref{fig:grmhd_var}, we show the variation in the principal axes FWHMs for a typical GRMHD simulation of the accretion flow of \sgra\ at 230\,GHz, both excluding and including the effects of scattering due to the interstellar medium in our line of sight \citep{Johnson_2016, Johnson_2018}. Although the simulation shows structural changes in the source morphology, deviations about the mean FWHM remain below 10\% for both the model and scattered simulation principal axes.

Furthermore, the emitting gas around supermassive black holes in low-luminosity active galactic nuclei becomes optically thin as we increase the observing frequency. The source extent is therefore dominated by the black hole shadow and Doppler-boosted features at higher frequencies~\citep{Falcke_2000}. This behavior is shown in Fig.~\ref{fig:grmhd_freq} for the GRMHD simulation of the quiescent accretion flow of \sgra\ observed at frequencies from 80\,GHz to 1\,THz \citep{Moscibrodzka_2018}. At frequencies of $\sim$300\,GHz and above, the source size changes very little with increasing frequency. These achromatic properties motivate the extrapolation of a source size from lower-frequency observations with short baselines, such as the EHT at 230\,GHz, to higher-frequency imaging experiments such as the upcoming EHT at 345\,GHz \citep{PaperII,Doeleman_astro2020}.

\begin{figure}[ht!]
\includegraphics[width=\linewidth]{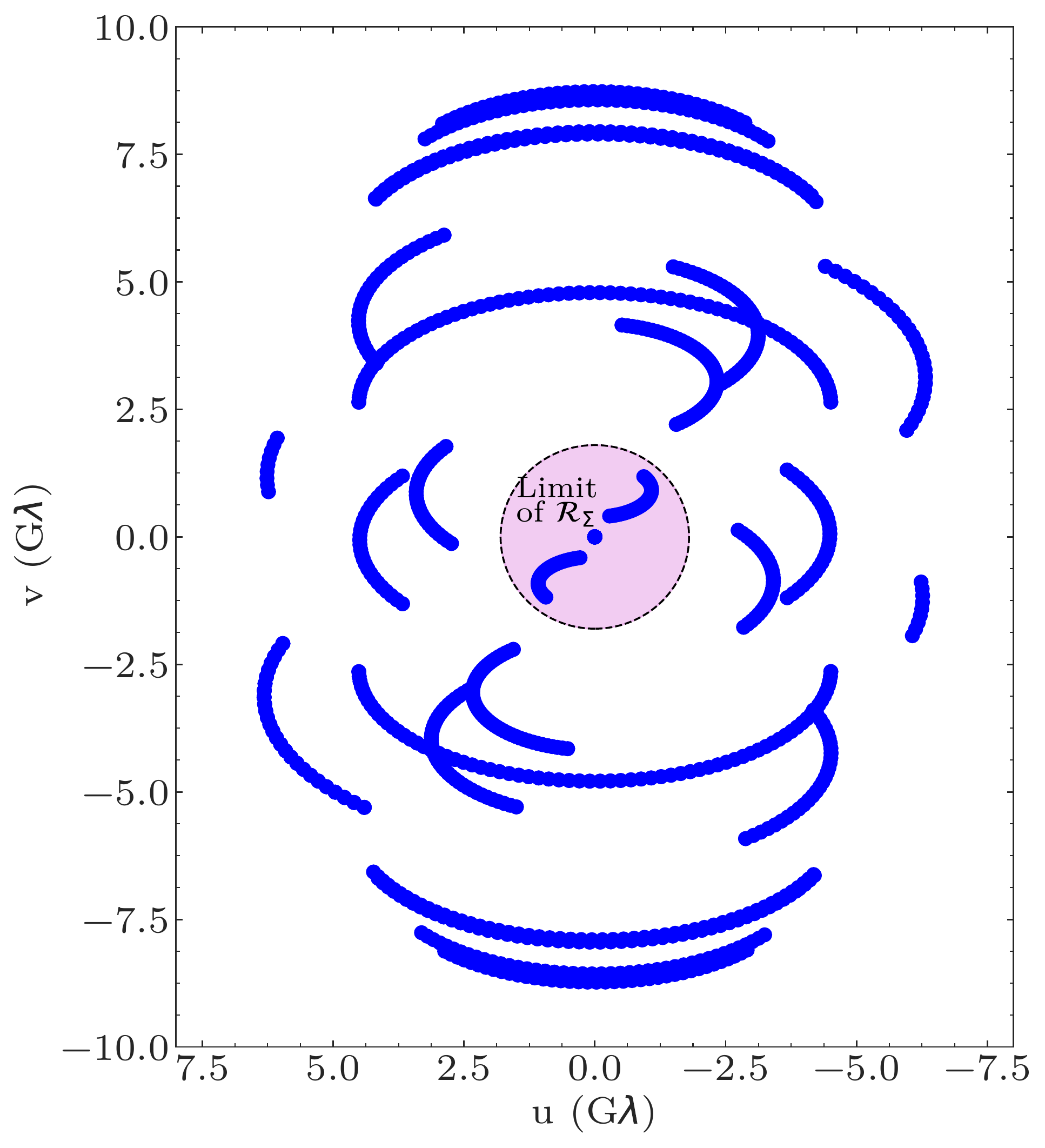}
\caption{$(u,v)$ coverage for simulated observations of \sgra\ with the EHT 2017 array at 230\,GHz. The magenta disk represents the range of $(u,v)$ constrained by the second moment regularization, with the boundary at the $1/e$ point of the corresponding visibility amplitude function for \sgra\ assuming an isotropic source of $60\,\mu$as FWHM \citep{Johnson_2018}.
}
\label{fig:coverage}
\includegraphics[width=\linewidth]{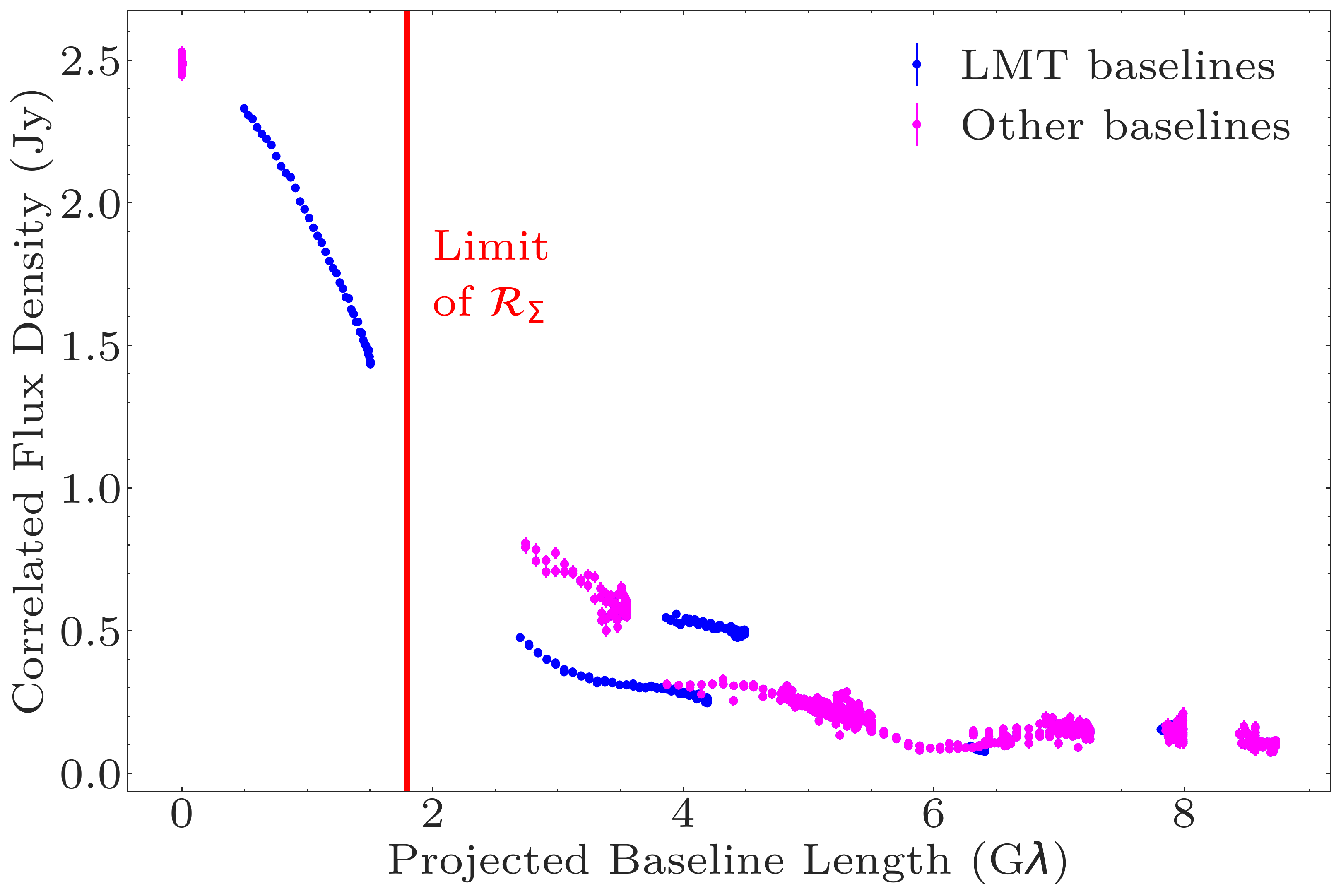}
 \caption{Visibility amplitudes for a model image of a semi-analytic advection-dominated accretion flow (ADAF) model of \sgra\ \citep{Broderick_2011} with a FWHM of $\sim60\,\mu$as as a function of $(u,v)$ distance sampled by the EHT in 2017 with and without the LMT (affecting mid-range baselines). The regularizer \rgauss\ governs the visibility amplitude behavior at short baselines until the $1/e$ point. This allows us to constrain and correct limitations and uncertainties in LMT calibration based on the expected behavior of the LMT--SMT mid-range baseline.} 
\label{fig:avery_radplot}
\end{figure}

\section{Demonstration}\label{sec:demo}
The second moment regularization can be used with informed size constraints from previous experiments, GRMHD simulations, or achromatic features from other observing frequencies. In this section, we demonstrate how the second moment regularization adds information to the imaging process if the data set to be imaged has no short baselines. For all following tests, we use a high $\beta_{\textrm{R}}=10^5$, such that the input source size is strongly constrained in the imaging process. To put this value into perspective, $\beta_{\textrm{R}}=10^5$ would cause a ${\sim}10\%$ difference between the input and reconstructed source sizes to be penalized equivalently to a change in reduced $\chi^2$ of ${\sim}1$ in our imaging procedure. This regularization weight tends to drive the second moment of reconstructed images to be within 20\% of the input values, therefore allowing some flexibility for the imaging process to deviate from the input second moment toward morphology favored by the available data. 

In Sect.~\ref{sec:demo1} we show improvements to the reconstructions when the source size is known. In Sect.~\ref{sec:demo2} we study the image quality and fidelity dependence on the assumed size in the regularization. Finally in Sect.~\ref{sec:demo3} we demonstrate that high fidelity images can be obtained without a priori knowledge of the source extent via input parameter searches.

\subsection{Imaging with complementary size constraints}\label{sec:demo1}
In Fig.~\ref{fig:coverage}, we illustrate the domain in which the second moment regularization (\rgauss) operates. The $(u,v)$ coverage is that of a typical observation of \sgra\ with the EHT at 230\,GHz. Assuming a source extent of 60\,$\mu$as from previous observations \citep{Johnson_2018}, the $1/e$ boundary of the visibility function for a source with that characteristic size is shown as a disk on the $(u,v)$ coverage. The only EHT baselines that lie within the \rgauss\ disk are intra-site baselines and the LMT--SMT short VLBI baseline. A single short VLBI baseline is very limited in constraining the overall extent of the source even assuming optimal performance of the telescopes. 

We selected a ray-traced image of a semi-analytic advection-dominated accretion flow (ADAF) model of \sgra\ \citep{Broderick_2011} with a similar characteristic size to the \sgra\ observations to assess the performance of the regularizer and to test the robustness of the imaging process as a function of the input parameters \xmaj, \xmin, and $\phi$. We sample the image with EHT 2017 coverage  (Fig.~\ref{fig:coverage}), where we have total flux density estimates from intra-site baselines and a valuable mid-range baseline (SMT--LMT) describing the extent of the source on the sky, as shown in Fig.~\ref{fig:avery_radplot}. We chose to discard all LMT baselines to limit the coverage and remove the constraining mid-range baseline for the regularizer tests. The extent of the source will then solely be enforced by the user-defined \xmaj, \xmin, and $\phi$ input parameters for \rgauss\ in the imaging process. It should be noted that imaging without the LMT not only removes short-baseline information on source extent but also long-baseline information on finer features, creating further differences in reconstructed images. The LMT, due to its size and central location, holds a strong weight in triggering decisions, while the SMT is a smaller and well-exercised station and is fairly flexible to various observing conditions. The choice to discard the LMT is thus mainly motivated by the known difficulties, to date, for the station to observe in a wide range of observing conditions and obtain adequate calibration information \citep{PaperIII,PaperIV}. Removing the SMT instead, for the purposes of these tests, would give similar results due to the lack of short-baseline information. 

\begin{figure*}[t]
\hspace{-0.04\linewidth}\raisebox{0.2\height}{\includegraphics[width=0.4\linewidth]{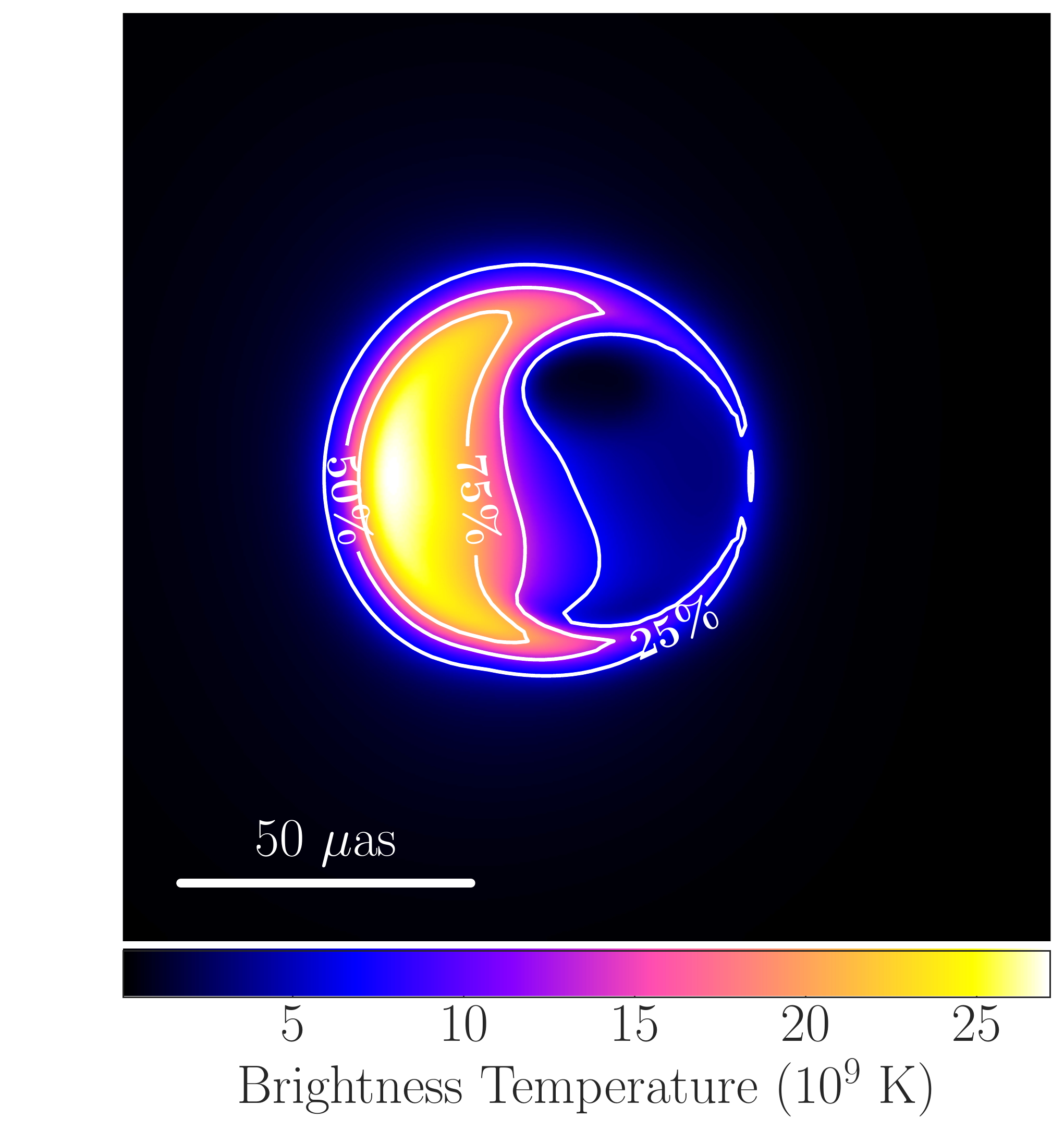}}\hspace{0.04\linewidth}
\includegraphics[width=0.59\linewidth]{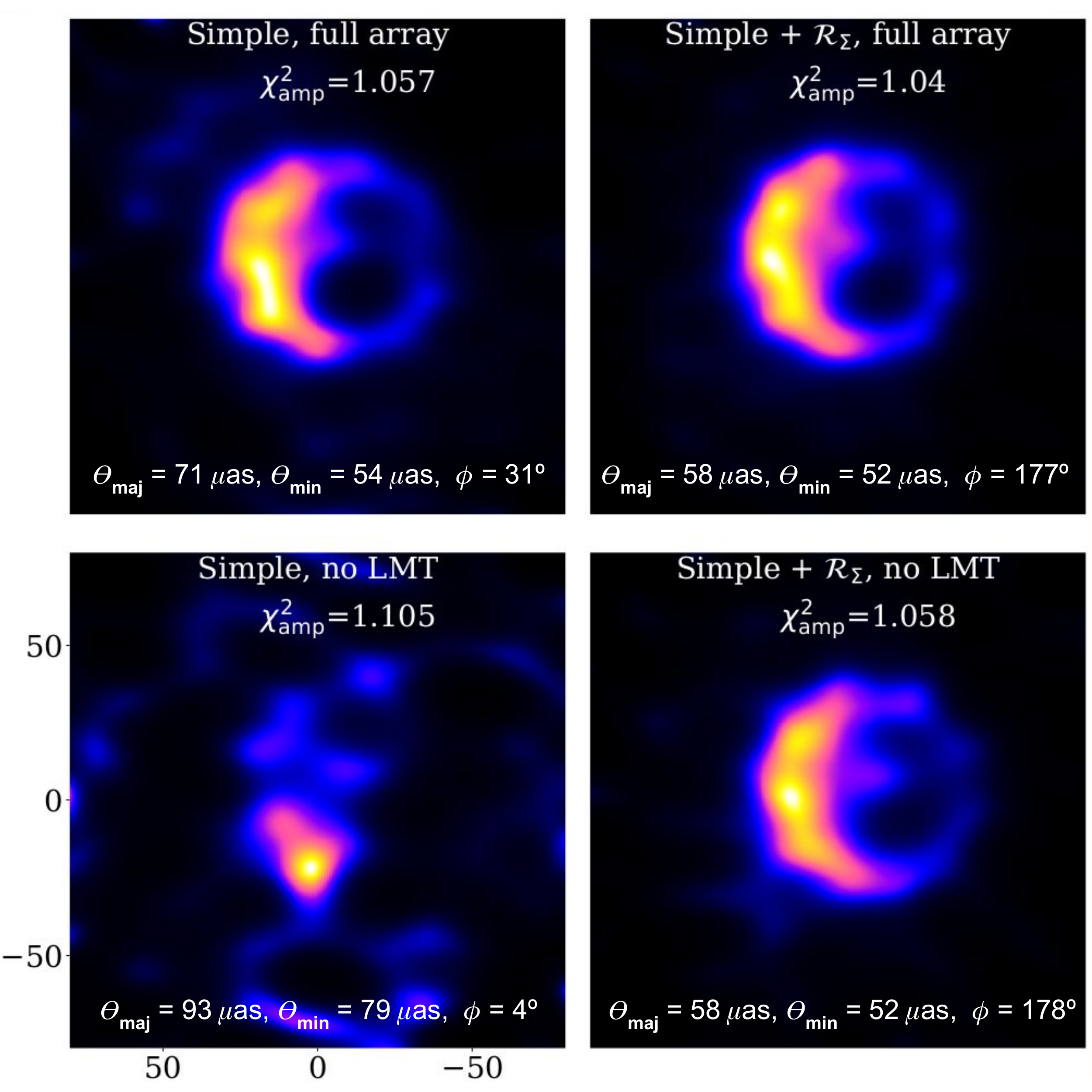}
\caption{{\it Left:} Model image of a semi-analytic ADAF model of \sgra \citep{Broderick_2011}, contours of 25, 50, and 75\% of the peak flux density are shown in white. {\it Right:} Tests of the second moment regularizer using the true image parameters as input ($\xmaj=58\,\mu$as, $\xmin=52\,\mu$as, $\phi= 177^\circ$ as measured directly from the model image), $\chi^2$ values are calculated for the data set without the LMT. We additionally give the resulting source size parameters for each reconstruction. Imaging of the example data set with full EHT 2017 coverage shows little difference between the imaging process with and without the second moment regularizer. When the LMT is removed, and thus the mid-range baseline no longer constrains the source size, \rgauss\ greatly improves the imaging. It should be noted that differences in finer features imaged with and without LMT are expected due to the loss of some long-baseline information from the removal of the LMT. 
}
\label{fig:avery_demo}
\end{figure*}

In Fig.~\ref{fig:avery_demo}, we show the model crescent image in the left panel, and example reconstructions for four different scenarios in the right panel. The first scenario is a reconstruction of the full EHT observations of the crescent, using closure quantities and visibility amplitudes, and maximizing simple image entropy. In that case, we obtain a good fit to the visibility amplitudes, and we recover an image very similar to the model image. Then, we reconstruct the same observations constraining the image to match the true second moment, as measured on the true image. With this method, we obtain a marginally improved fit to the amplitudes, but visibly less diffuse flux outside the crescent due to the constraint of \rgauss. Once we remove the LMT however, the simple imaging with maximum entropy is not able to reconstruct the morphology of the source, although some compact features are reconstructed that enable a decent fit to the visibility amplitudes. When adding \rgauss\ to the process, the second moment constraint is able to offset the absence of short baselines and reconstructs an image of improved quality in terms of both image morphology and goodness-of-fit to the amplitudes. This demonstration shows that \rgauss\ successfully adds additional information to reconstruct a more physically plausible image even when mid-range baselines are lacking in the underlying data set. The improvement in the amplitude $\chi^2$ also shows that \rgauss\ is a useful tool to aid convergence in imaging.

\begin{figure}[t]
\centering 
\includegraphics[width=0.98\linewidth]{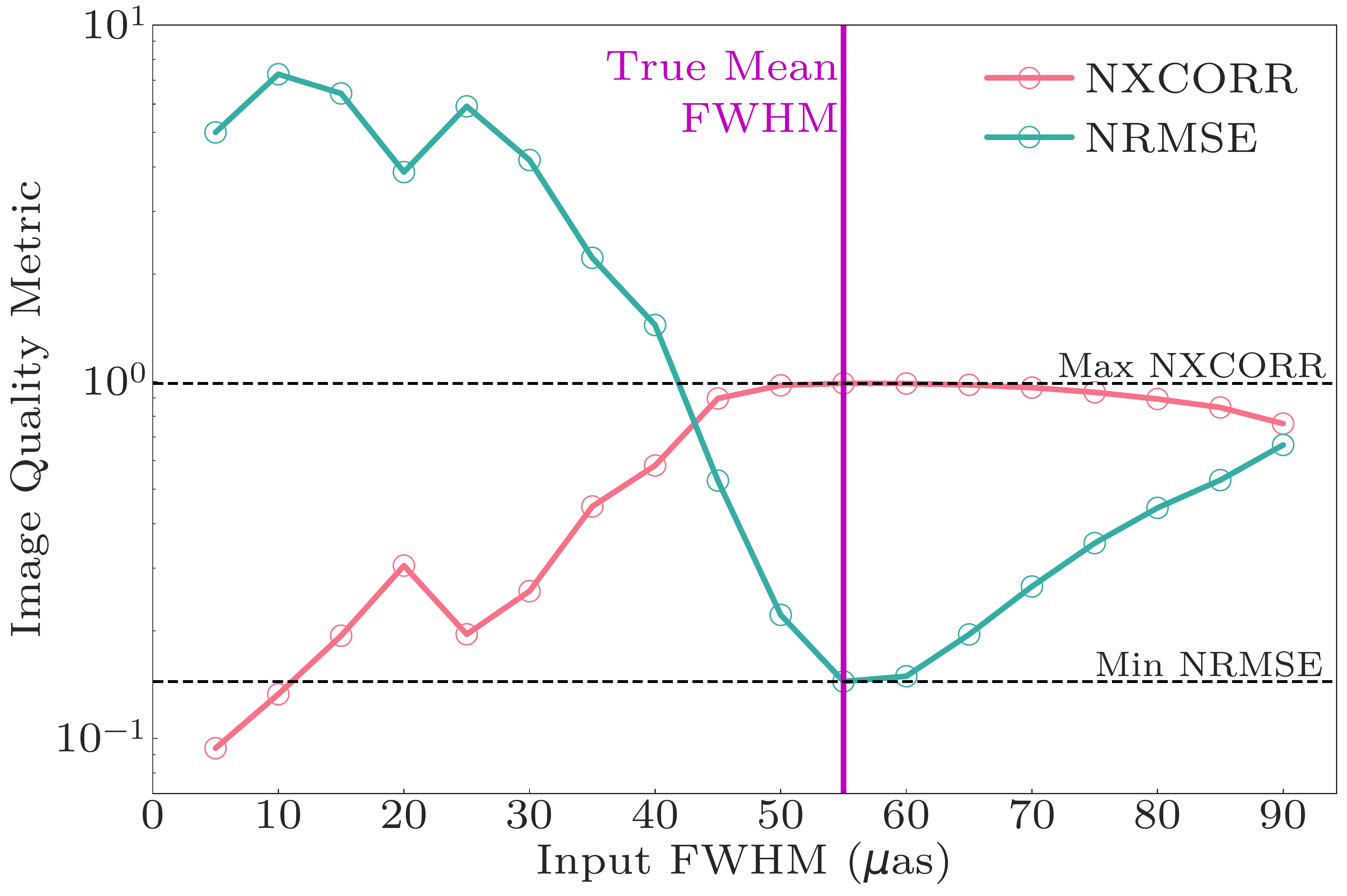}
\includegraphics[width=0.98\linewidth]{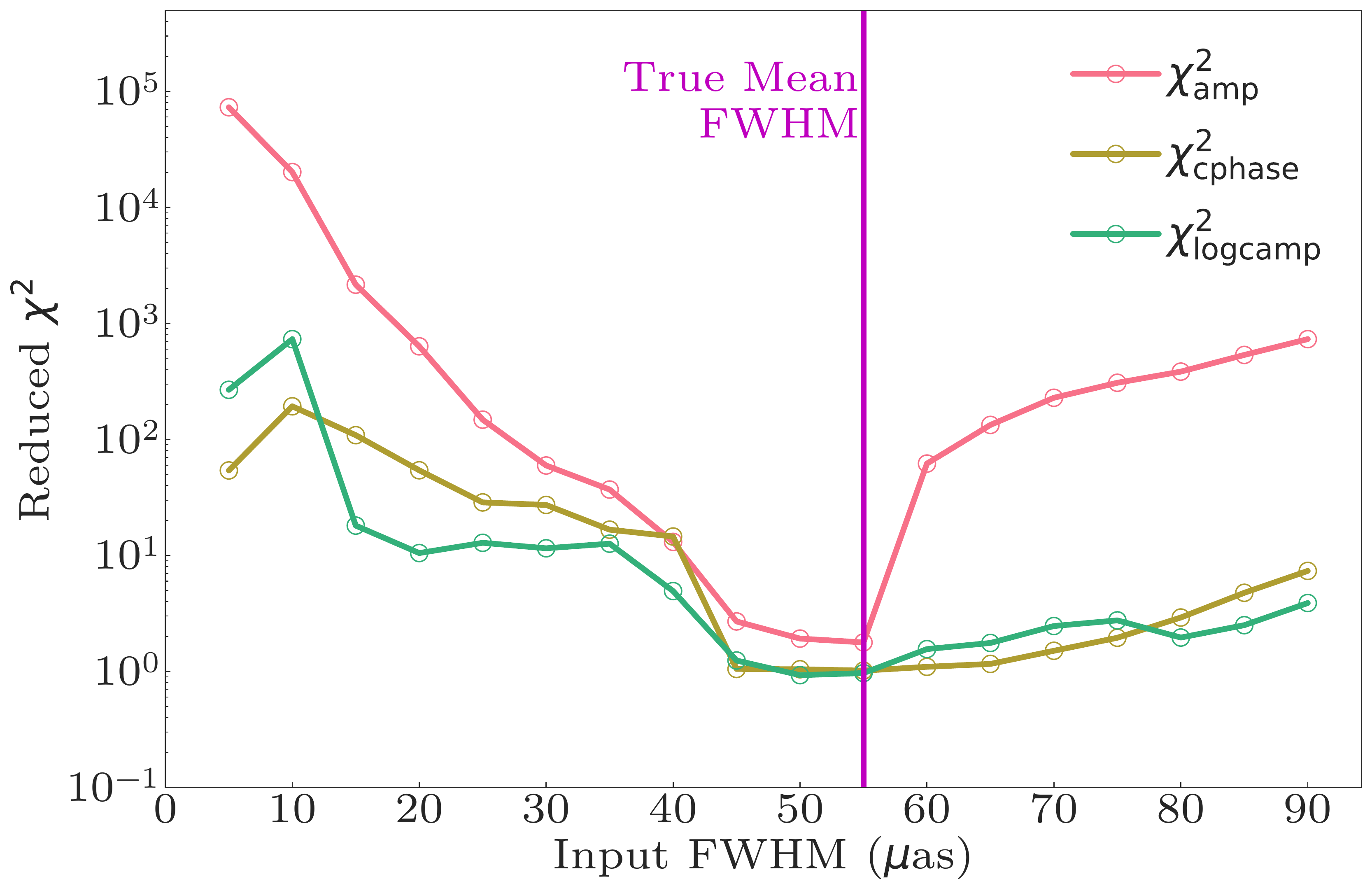}
\caption{Quality of the images obtained with different input FWHM (major and minor axes equal, position angle is zero). The image quality is measured in three ways: (1) the normalized cross-correlation against the true image, or NXCORR; (2) the normalized root-mean-square error against the true image, or NRMSE, shown in the top panel; and (3) reduced $\chi^2$ goodness-of-fits to the three data products used in the reconstructions (visibility amplitudes, closure amplitudes and phases) shown in the bottom panel. NRMSE is more sensitive to subtle differences in the images than NXCORR due to the higher weight associated with large pixel-by-pixel errors and is minimized in a comparable range of input FWHMs to the reduced data $\chi^2$. The narrow range of FWHMs encompasses the true mean source FWHM (magenta vertical line). 
}
\label{fig:avery_metrics}
\end{figure}

\subsection{Dependence of reconstructed images on assumed size}\label{sec:demo2}
In the demonstration of \rgauss\ we constrained the second moment to the true size of the source, to enable an accurate reconstruction of the image. However, in practice, the true size of the source is unknown, and is instead approximated from Gaussian model fitting to closure quantities and/or short-baseline visibility amplitudes and extrapolated from historical measurements. We therefore investigate the robustness of the image reconstructions when the input Gaussian parameters are strongly enforced in the imaging process, corresponding to a strong weight of the \rgauss\ hyperparameter, while changing input principal axes FWHMs. We demonstrate this dependence by imaging the data set of the crescent model sampled by the EHT 2017 coverage without the LMT, such that the extent of the source is only enforced by the varying inputs to \rgauss. For simplicity, we use a single common imaging script varying only the input principal axes FWHMs. We assume an isotropic source size such that \xmaj = \xmin\ and $\phi=0^\circ$, and a range of input FWHMs of $5-90\,\mu$as. 

We utilize two metrics to compare the quality of the reconstructed image to the true model image. The normalized root-mean-square error (NRMSE) of each image is given by:
\begin{align}
    {\rm NRMSE} = \sqrt{\frac{\sumk (\Ik - \Ik')^2}{\sumk \Ik^2}}\, , 
\end{align}
where $I'$ is the intensity of the reconstructed image and $I$ is that of the true image~\citep[e.g.,][]{Chael_2018}. If the reconstructed image is identical to the true image, the NRMSE is zero. Therefore, the input FWHM for the reconstruction resulting in the minimum NRMSE in comparison to the true image gives the best fit. 

The normalized cross-correlation (NXCORR) is a sliding inner-product of two normalized functions. For fast numerical computation, we determine the cross-correlation of the Fourier transforms of the normalized intensity patterns of the true image $I_\mathrm{norm}$ and the reconstructed image $I'_\mathrm{norm}$ at different relative shifts $\boldsymbol{\delta}$ across the extent of the images. For each pixel i in the image, we normalize the intensity via:
\begin{align}
    I_{\mathrm{norm,i}} = \frac{\Ik - \mu_I}{\sigma_I}\, , 
\end{align}
where $\mu_I$ and $\sigma_I$ are the mean and standard deviation of the intensity distribution in the image. The cross-correlation for a given shift $\boldsymbol{\delta}$ is then given by:
\begin{align}
{\rm NXCORR}(\boldsymbol{\delta}) = |\mathcal{F}^{-1}\{\mathcal{F}\{I_\mathrm{norm}^\ast(\mathbf{x})\} \cdot \mathcal{F}\{I_\mathrm{norm}'(\mathbf{x}+\boldsymbol{\delta})\}\}|.
\end{align}
The shift at which the cross-correlation is maximized is then used to output the final NXCORR value for the two images. This method is less sensitive to individual features in the reconstructed image than NRMSE as it compares the bulks of each intensity pattern as opposed to the NRMSE pixel-to-pixel comparison. The $\chi^2$ statistics follow the equations presented in Sect. 2.1 of \citet{PaperIV}. 

\begin{figure}[t]
\centering 
\includegraphics[width=\linewidth]{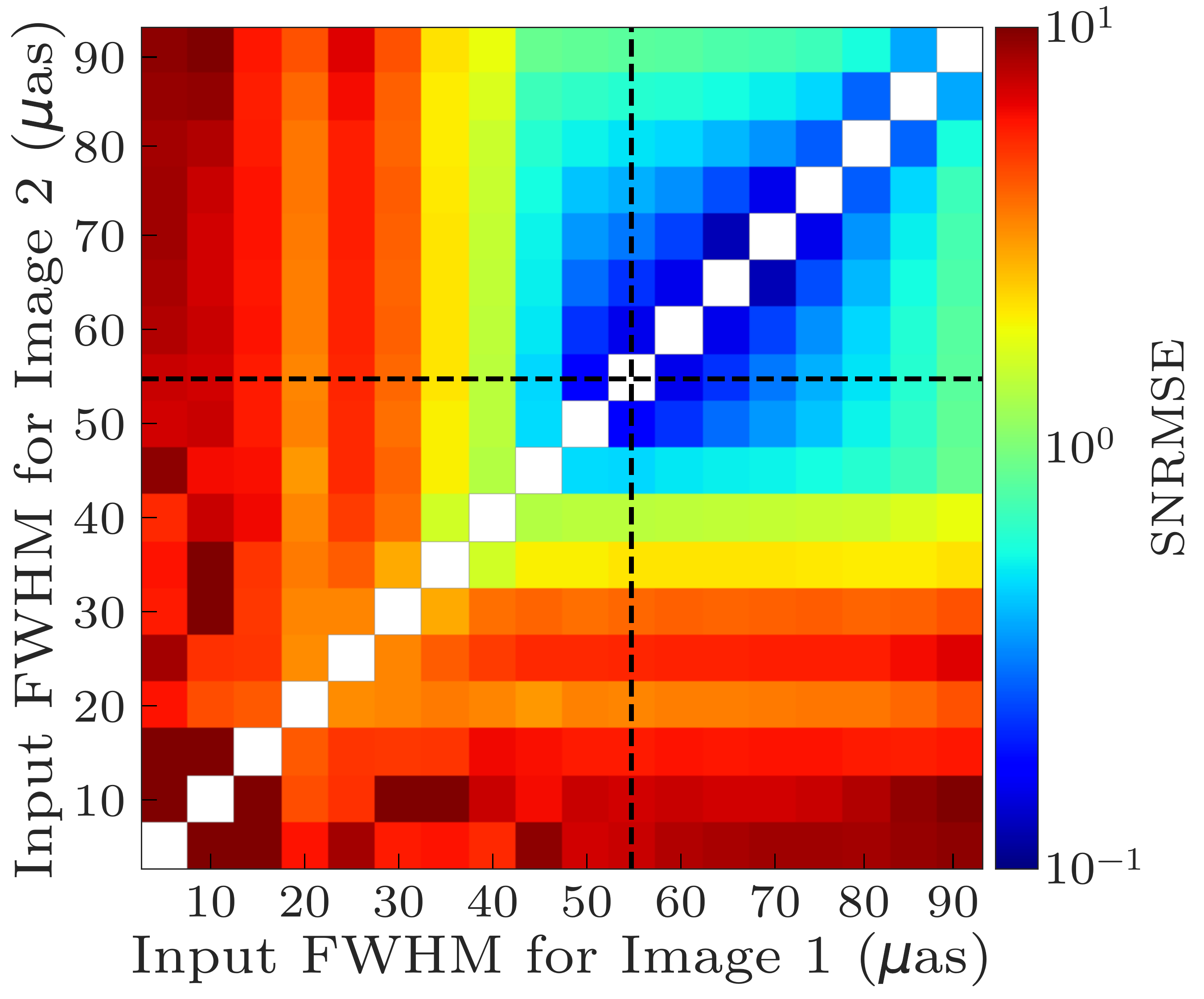}
\caption{Cross-comparisons of reconstructed images with varying isotropic input FWHMs using symmetrically normalized root-mean-square error (SNRMSE). The SNRMSE grid shows a greater correspondence of images with input FWHMs near the true mean FWHM of $55\,\mu$as, marked by the dashed black lines. The reconstructed images with varying input size (5--90\,$\mu$as) are all compared to each other, where image 1 and image 2 are the two images to be compared ($I'_1$ and $I'_2$ respectively in Equation~\ref{eq:snrmse}). The diagonal is each image compared to itself. The SNRMSE grid gives a range of plausible input FWHMs for \rgauss\ that result in high fidelity images even when the true source size is unknown.}
\label{fig:nrmse_avery}
\end{figure}

In Fig.~\ref{fig:avery_metrics}, we show the NRMSE and NXCORR metric scores for the reconstructed images compared against the true image (left panel of Fig.~\ref{fig:avery_demo}), and the reduced data $\chi^2$ goodness-of-fits to the imaged data set (Fig.~\ref{fig:avery_radplot}, no LMT). The NXCORR is maximized at an input FWHM of $55\,\mu$as, and the NRMSE is minimized at the same input FWHM. This value corresponds to the mean FWHM (average of \xmaj = $58\,\mu$as and \xmin = $52\,\mu$as) of the true image. With this test, we find an excellent correspondence between the reconstructed image with the highest quality (highest NXCORR, lowest NRMSE, and lowest reduced data $\chi^2$) and the image with the input \rgauss\ FWHM closest to the true value. Images with input FWHMs close to the optimal value are of similarly good quality. We thus show a good performance of \rgauss\ in the imaging process even with input sizes inaccurate to within 20\% of the true size. The reduction in data $\chi^2$ values as we approach the true source size also indicates that \rgauss\ gives a convergence boost toward a higher fidelity image. This behavior is caused by \rgauss\ rapidly reducing the favored set of images to only those that constrain flux within a given region. The region limits that best represent the flux distribution in the true image allow the minimizing process to focus more quickly on the data terms and achieve better reduced $\chi^2$ values within the given imaging conditions. This property also allows us to survey the response of the imaging process and goodness-of-fits to the available data via parameter searches over different favored second moments (and thus favored flux regions) and determine optimal parameters that best represent properties of the data set.

\subsection{Imaging without complementary size constraints}\label{sec:demo3}
The NRMSE metric proves to be more sensitive to differences in the image structure than NXCORR, as shown in Fig.~\ref{fig:avery_metrics}, due to the higher weight associated with large errors in the computation of the NRMSE. For that reason, we have selected NRMSE to score comparisons between the reconstructed images themselves. For this test, we assume that the true image and true FWHM are unknown, as is the case for real experiments. We instead focus on the morphological characteristics that appear in the images based on the underlying data, and how the inputs to \rgauss\ affect the correspondence between reconstructed images. We restructure the metric into a symmetrically-normalized root-mean-square error \citep[SNRMSE;][]{hanna1985,Mentaschi_2013} to render the NRMSE independent of the input and comparison image choice:
\begin{align}
    {\rm SNRMSE} = \sqrt{\frac{\sumk (I'_{1,i} - I'_{2,i})^2}{\sumk I'_{1,i} I'_{2,i}}}. \label{eq:snrmse}
\end{align}
Here $I'_1$ and $I'_2$ are the two reconstructed images to be compared. 
In Fig.~\ref{fig:nrmse_avery}, we show an SNRMSE grid comparing each reconstructed image to all others, where the diagonal squares correspond to each image compared with itself. We have marked with dashed lines where the mean FWHM of the true image lies. We find that images with input FWHMs near the true FWHM of the source have a better SNRMSE with each other than all other combinations of images. This test enables the user to find a range of characteristic sizes minimizing SNRMSE via a size parameter search. For compact sources that are distinctly elliptical, a one-dimensional size parameter search is useful to quickly sweep through a wide range of sizes and determine a range of plausible sizes for the source extent. A search within that range, varying parameters in two dimensions ($\xmaj$, $\xmin$, and $\phi$), can then be carried out to refine the source size estimate for the imaging process.

We find that the use of the regularizer improves the quality of the resulting image even if the input parameters deviate by 20\% from the true values. We also find that the strong use of the regularization, when combined with a size parameter search, is able to converge toward the true FWHM values, even when the true source dimensions are unknown. The use of SNRMSE and $\chi^2$ statistics serve well to score individual images and parameters without a priori knowledge of the source extent. 

\begin{figure*}[t]
    \centering
    \includegraphics[width=0.33\linewidth]{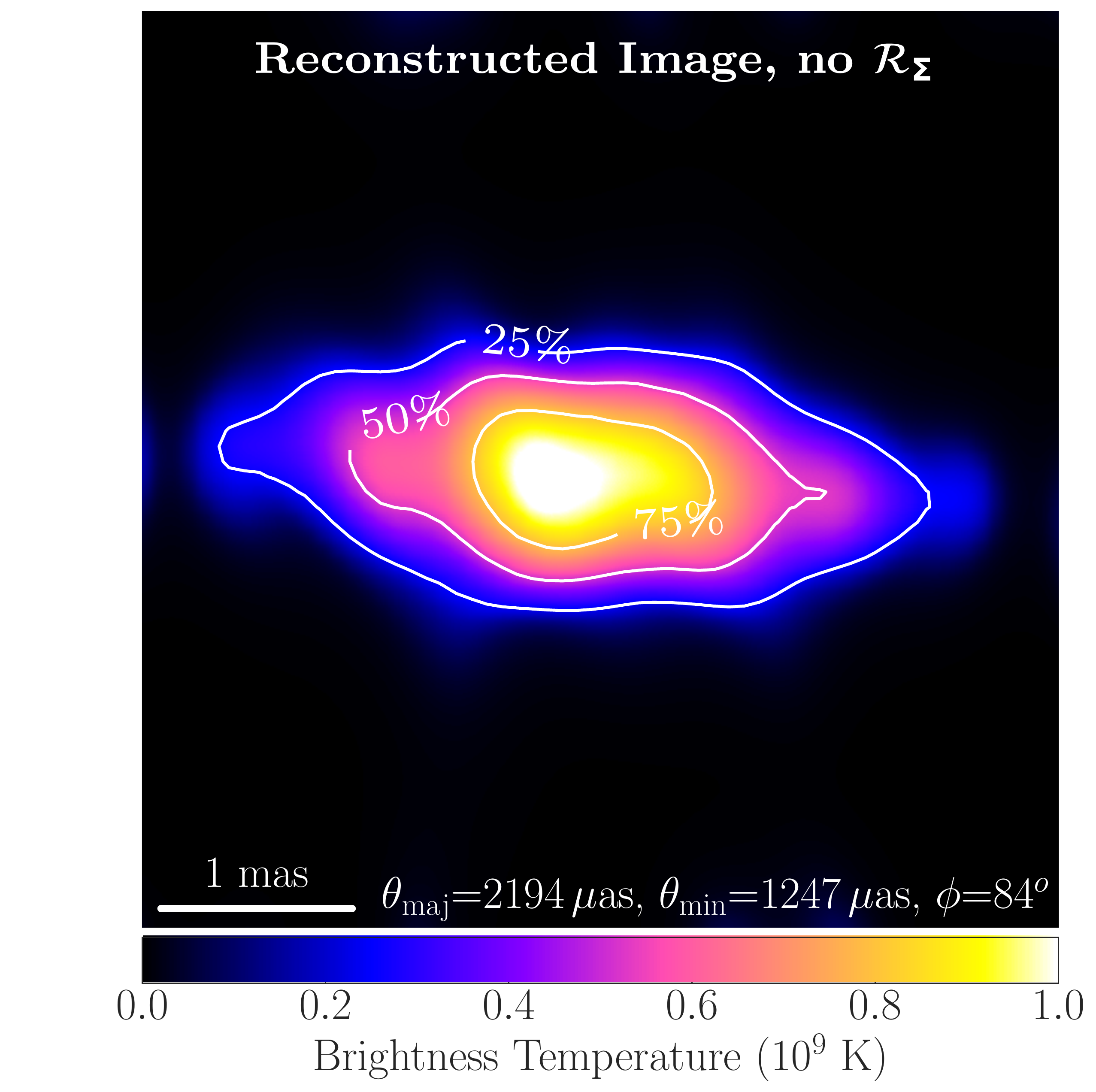}\hspace{-0.01\linewidth}
    \includegraphics[width=0.33\linewidth]{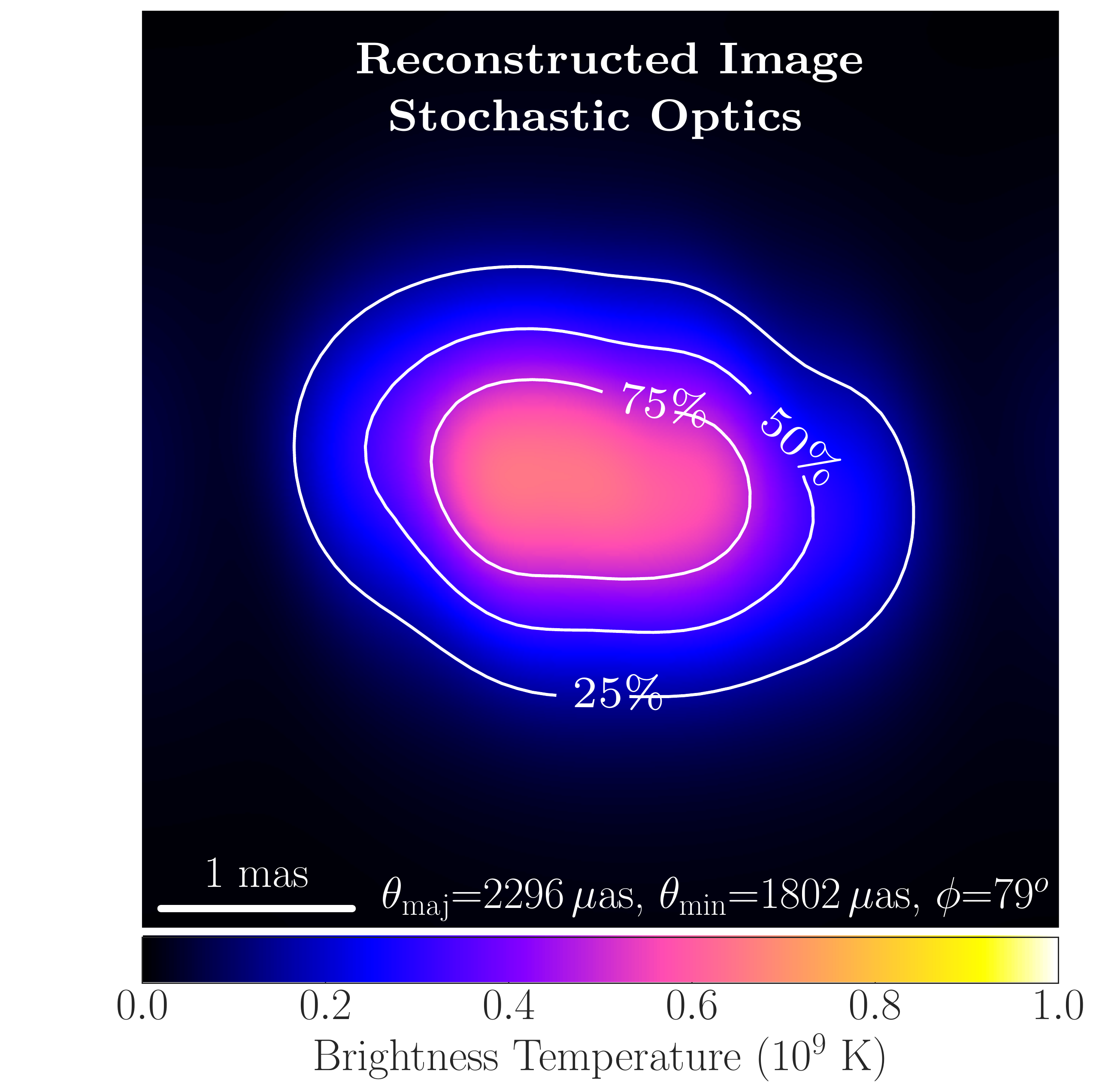}\hspace{-0.01\linewidth}
    \includegraphics[width=0.33\linewidth]{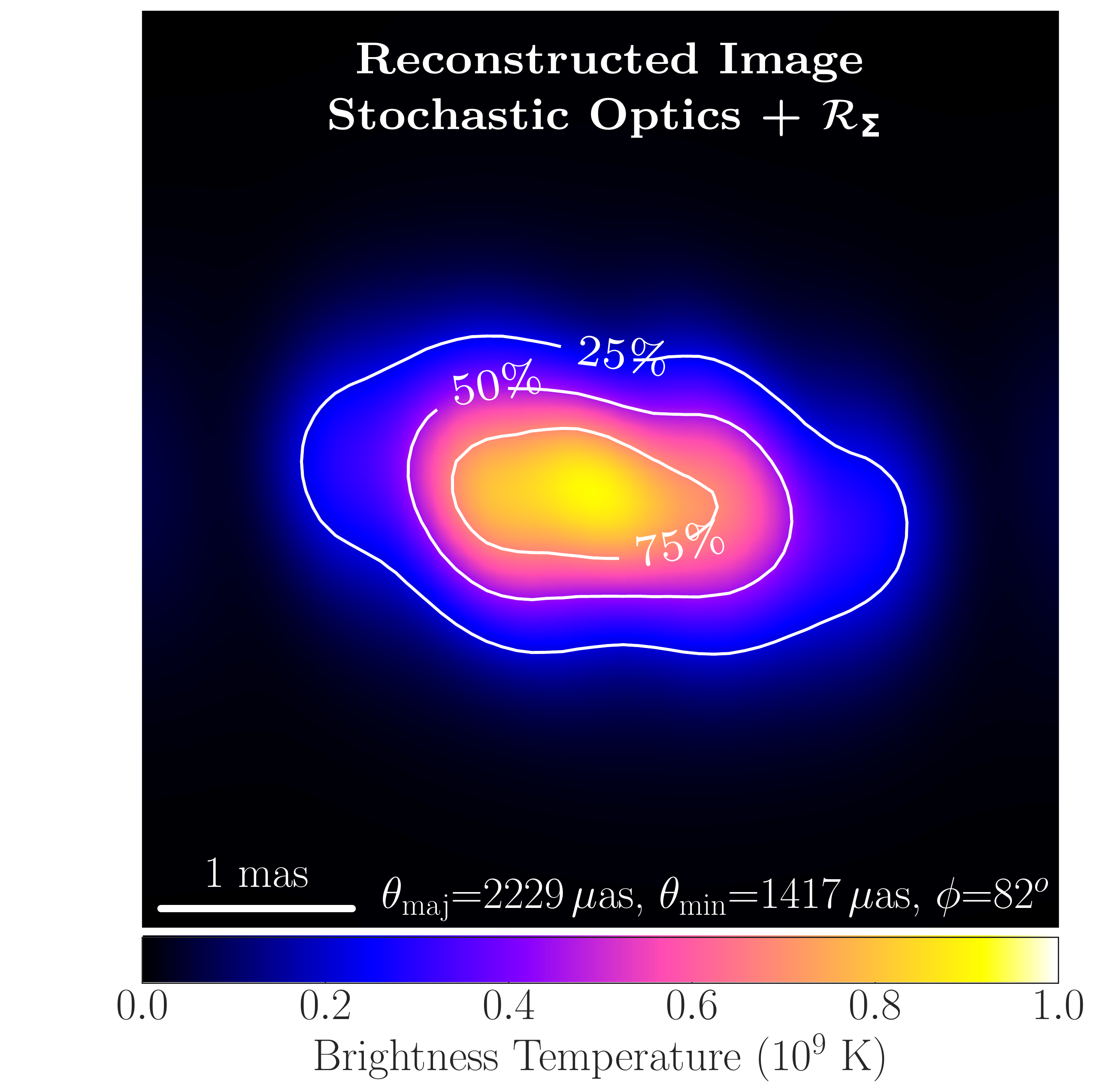}
    \caption{Reconstructions of 22\,GHz VLBA+GBT observations and their resulting source extents. MK and SC have no detections, and HN and NL are flagged due to their very low sensitivity in this experiment. {\it Left}: a simple reconstruction of the scattered image without \rgauss.
    {\it Center}: a reconstruction of the scattered image via stochastic optics \citep{Johnson_2016}, using the scattering model by \citealt{Johnson_2018}. {\it Right}: a reconstruction with stochastic optics, using \rgauss\ and the input source size as determined by \citealt{Johnson_2018} from high-precision Gaussian model fitting: $\xmaj = 2255 \pm 61\,\mu$as, $\xmin = 1243 \pm 39\,\mu$as, $\phi = 81.9 \pm 0.2\degree$.
    The reconstruction with \rgauss\ helps constrain the extent of the source in the north--south direction, where measurements are lacking due to the predominantly east--west configuration of the VLBA+GBT. 
    }
    \label{fig:22ghz_im}
\end{figure*}

\section{Applications}\label{sec:examples}
In addition to simple static imaging, second moment regularization can easily be coupled to more sophisticated and complex imaging techniques. In Sect.~\ref{sec:stochastic} we present an example of the use of second moment regularization for scattering mitigation imaging of \sgra\ at longer wavelengths. In Sect.~\ref{sec:movie} we demonstrate how second moment regularization in individual sparse snapshots improves the quality of dynamical reconstructions of variable sources, such as a movie of an orbiting "hot spot" in \sgra's accretion flow.

\subsection{Scattering mitigation}\label{sec:stochastic}

The second moment constraint in imaging can both be used for data sets where short baselines are lacking, as demonstrated in Sect.~\ref{sec:demo}, and for data sets where short-baseline measurements have large uncertainties due to difficult observing conditions. An example of the latter case is presented in \citet{Issaoun_2019}, where observations of \sgra\ at 86\,GHz with the Global Millimeter VLBI Array and ALMA (project code MB007) yielded high signal-to-noise (SNR) detections on long baselines but bad weather at select Very Long Baseline Array (VLBA) stations led to poorly constrained short-baseline measurements. Imaging of the source with RML would not have been feasible with these measurements alone, as the large uncertainties in the short-baseline measurements caused flux to spread nonphysically across the reconstructed images. Since the size of \sgra\ on the sky is well studied and known to be affected by anisotropic scatter-broadening from the interstellar medium~\citep{Davies_1976,vanLangevelde_1992,Frail_1994,Bower_2004,Shen_2005,Bower_2006,Psaltis_2018,Johnson_2018}, previous size measurements \citep{Ortiz_2016,Brinkerink_2019} were used to constrain the extent of \sgra\ in the imaging process with \rgauss. In this manner, we obtained an image that was able to fit new long-baseline detections to ALMA, likely refractive noise from scattering substructure. 

The second moment regularization was also implemented in the scattering mitigation code {\em stochastic optics} developed by \citet{Johnson_2016}. Stochastic optics aims to mitigate the effects of scattering to derive an intrinsic (unscattered) image of the source. The code solves for the unscattered image by separating and mitigating the two main components of the \sgra\ scattering screen: the diffractive scattering that causes the image to become a convolution of the true image and the scattering kernel; and the refractive scattering that introduces stochastic ripples that further distort the image. The stochastic optics framework therefore simultaneously solves for the unscattered image and the scattering screen assuming a given model for the diffractive blurring kernel and the time-averaged refractive properties. The model assumed here is the \citet{Johnson_2018} scattering model, the best-fitting model to \sgra\ observations to date \citep{Issaoun_2019}.

The implementation of \rgauss\ in stochastic optics only constrains the size of the scattered source (\sgra\ as we see it on the sky) based on historical measurements from model fitting, such that the technique can more accurately mitigate the effects of interstellar scattering to obtain a physically motivated intrinsic image of the accretion flow of \sgra\ \citep[for further details, see][]{Issaoun_2019}. The intrinsic image itself is not directly constrained by the second moment regularization, but is derived from the combination of the constrained scattered image and knowledge of the interstellar scattering.

Here we illustrate the use of \rgauss\ within stochastic optics using a lower frequency data set. Observations of \sgra\ at 22\,GHz with the VLBA+GBT (project code BG221A) showed clear long-baseline detections of refractive noise from interstellar scattering \citep{Gwinn_2014,Johnson_2018}. These long-baseline detections should translate to substructure in the image, distorting the intensity pattern seen for \sgra\ away from the scatter-broadened smooth elongated Gaussian-like morphology. While the scattering substructure is very apparent in the data set, it is a non-trivial task to successfully show its effects on the image itself and obtain an intrinsic image of the source. This is due to the imaging process being driven predominantly by the abundance of intra-VLBA short-baseline measurements in comparison to the few VLBA--GBT long-baseline detections. We therefore test the addition of \rgauss\ on this data set, using the source dimensions in Table 1 of \citet{Johnson_2018} from elliptical Gaussian model fitting.

In Fig.~\ref{fig:22ghz_im}, we show three separate reconstructions of the 22\,GHz data set. A standard RML reconstruction of the data set (Fig.~\ref{fig:22ghz_im} left panel) shows some distortions in the scattered image, but the morphology remains fairly smooth and elongated. Standard RML imaging cannot solve for the scattering properties, therefore the procedure is solely focused on obtaining the highest fidelity scattered image possible from the data set. We will thus treat this image as our comparison image for this data set.  
When using stochastic optics however, the imaging process is more complex, as it is simultaneously imaging the scattered source and solving for the scattering properties to disentangle scattering from intrinsic source structure. This process derives a scattered image that is not well-constrained in the north--south direction due to the configuration of the VLBA+GBT, resulting in a large source image that is not fully converged to the image obtained from standard RML (Fig.~\ref{fig:22ghz_im} center panel). Since the scattered image does not match our expectations of the physical morphology of the source, the derived intrinsic image should also not be trusted. The challenge is then to improve the convergence of the imaging component of stochastic optics to quickly obtain a physically motivated scattered image and therefore undergo a higher-fidelity separation of the scattering and intrinsic structure. When using \rgauss, where the scattered image is constrained to remain within the size obtained by \citet{Johnson_2018} using elliptical Gaussian model fitting, the resulting scattered image is more elongated in the east--west direction (Fig.~\ref{fig:22ghz_im} right panel) and showing distortions similar to those of the standard RML reconstruction  This shows that the use of \rgauss\ helps the convergence of the scattered image through stochastic optics to a more physically motivated reconstruction, and thus will give a more realistic underlying unscattered image of the source. 

\begin{figure*}[t]
\centering 
\includegraphics[width=1.\linewidth]{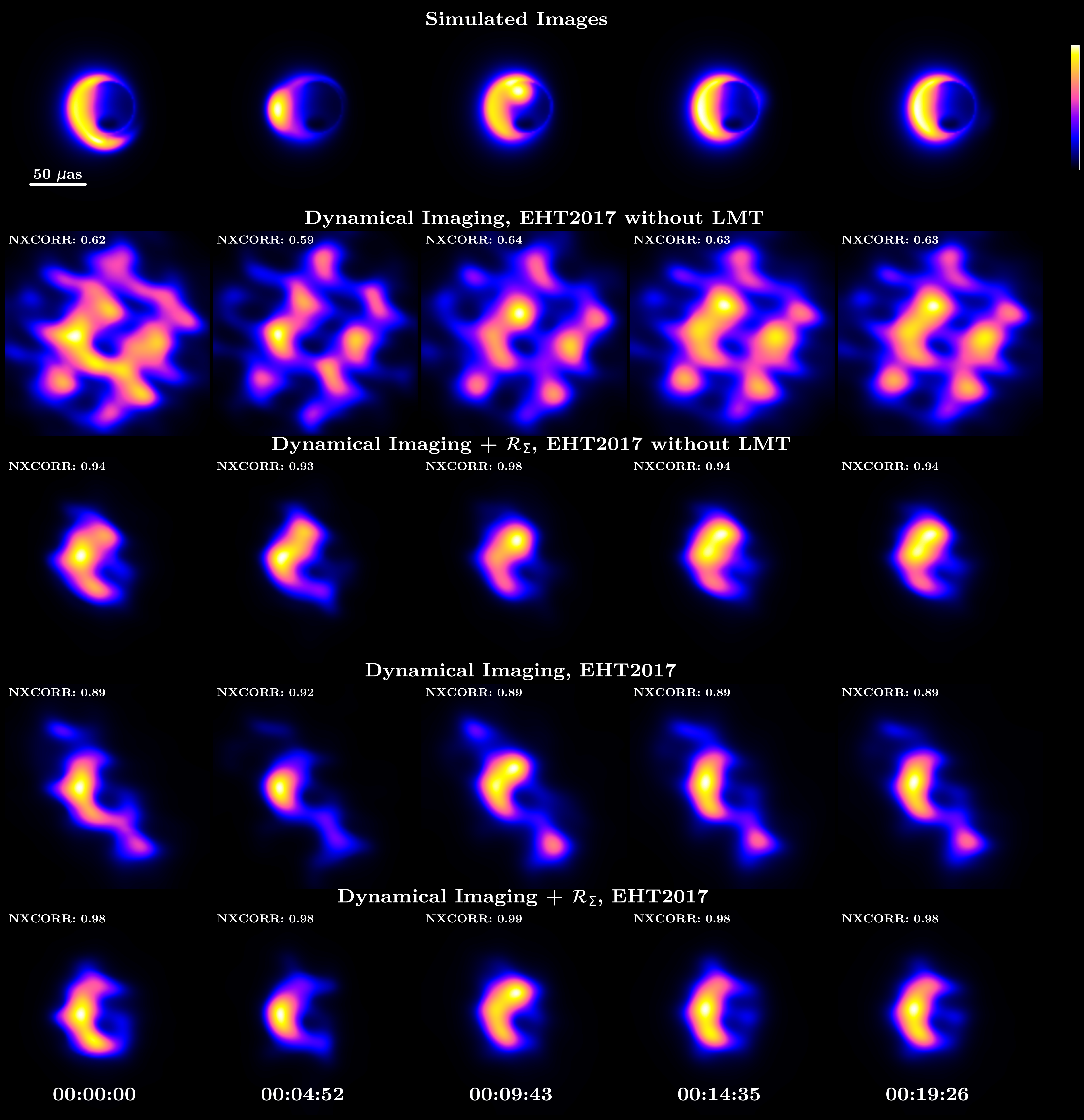}
\caption{Reconstruction of a simulated flare using dynamical imaging \citep{Johnson_2017}. From top to bottom: simulated images of a flare with a period of 27 minutes (model B of \citealt{Doeleman_2009b}); simple dynamical imaging without the LMT (no short-baseline points constraining the source extent); dynamical imaging using \rgauss\ without the LMT (the second moment regularization offsets the lack of short-baseline constraints); simple dynamical imaging with full EHT2017 sampling; dynamical imaging using \rgauss\ with EHT2017 sampling. Using \rgauss\ significantly improved the quality of dynamical reconstructions both with the full array and without the LMT. NXCORR values against the model images are shown in the top left corner for each reconstructed snapshot. The variations in the resulting FWHMs of the reconstructed images are visually evident.}
\label{fig:hotpspot}
\end{figure*}

\subsection{Dynamical imaging}\label{sec:movie}
There are additional applications for the second moment regularization in movie reconstructions of variable sources where single snapshots have very sparse coverage. We can test the robustness of movie reconstructions with the loss of short baselines using a simulated flare \citep[model B of][]{Doeleman_2009b} with an orbiting period of 27\,minutes around the same crescent model as in Sect.~\ref{sec:demo}. We reconstruct movies of the orbiting ``hot spot'' using {\em dynamical imaging}, enforcing temporal continuity between individual frames \citep[for further details, see][]{Johnson_2017}. We reconstruct a movie of the orbiting hot spot for four different scenarios: (1) we use the EHT 2017 array without the LMT, no short baselines are present in the individual snapshots to constrain the source extent; (2) we use the data set without the LMT, but constrain the extent of the source (the dimensions of the crescent model) with \rgauss, (3) we use the full EHT 2017 to reconstruct the orbit; (4) we use the full EHT 2017 and \rgauss\ to reconstruct the orbit. In Fig.~\ref{fig:hotpspot}, we show individual frames of the true simulation and of the reconstructed movies for the four different scenarios. The reconstructions without \rgauss\ either yield unphysical source structure dominated by the dirty image (due to the lack of information without LMT) or contain imaging artifacts from flux spreading due to the sparse coverage of individual snapshots. In particular, even with the full EHT2017 array, dynamical imaging without \rgauss\ shows north--east and south--west artifacts from the dirty image that persist due to the sparse snapshot coverage. The reconstructions with \rgauss, even without the LMT, are significantly cleaner and more accurately reconstruct the motion and morphology of the simulation, as shown by NXCORR results when compared to the truth simulated images.

\section{Summary}\label{sec:summary}

In summary, we have developed a regularization function \rgauss, for use in a regularized maximum likelihood framework for interferometric imaging, that constrains the spread of flux in reconstructed images to match input parameters defined by the user. The second moment regularization is a natural extension of common imaging tools, such as image total flux and image centroid constraints (zeroth and first moment respectively), that help to mitigate the missing information problem in high frequency VLBI. The regularization assumes that the source is compact, with a stable size, and is resolved on longer baselines of the interferometer. The validity of these assumptions for the EHT's primary targets, \sgra\ and M87, are well-motivated by state-of-the-art GRMHD simulations and long-term observational studies. For well-studied sources, this method allows for contingency against weather, a major deterrent for high frequency VLBI, and gives more flexibility for triggering decisions if key short baselines yield poorly constrained measurements or become unavailable during or between observations. 

We have shown that \rgauss\ successfully informs the source behavior on short baselines and is defined only by three Gaussian parameters and the regularization hyperparameter. Imaging with \rgauss\ is able to reconstruct high fidelity images fitting to the data products even if the input source dimensions deviate from the true values by up to 20\%. The regularization therefore gives a larger flexibility than needed to account for changes in size from, for example, GRMHD simulations of highly variable sources such as \sgra. We have also shown that parameter searches over a range of isotropic FWHMs using \rgauss\ in conjunction with goodness-of-fit statistics to data products and symmetrically-normalized root-mean-square error of image comparisons help determine high-fidelity source extent even if the exact size and morphology are unknown. 

The regularization can be used to image with any choice of data products and any choice of feature-driven regularizers within the framework of the {\tt eht-imaging} library \citep{Chael_2016,Chael_2018} and is easily transferable to other tools or other RML imaging packages \citep[e.g., \texttt{SMILI};][]{Akiyama_2017b,Akiyama_2017a}. We have shown that the \rgauss\ implementation complements other techniques tackling source properties that add difficulty and complexity to the imaging process, such as time variability \citep[via dynamical imaging;][]{Johnson_2017,Bouman_2017} and interstellar scattering \citep{Johnson_2016,Issaoun_2019}. Source parameter inputs can either be obtained from model fitting to abundant short-baseline measurements, historical measurements from observations with short baselines present, extrapolated from other wavelengths based on achromatic features, or estimated via parameter searches. The second moment regularization could prove particularly useful in future work with the EHT, both for dynamical reconstructions of variable sources such as \sgra\ and for upcoming imaging observations at 345\,GHz \citep{PaperII,Doeleman_astro2020}.

\begin{acknowledgements}
We thank John Wardle for his helpful comments and careful review. This work is supported by the ERC Synergy Grant ``BlackHoleCam: Imaging the Event Horizon of Black Holes'', Grant 610058. We thank the National Science Foundation (AST-1440254, AST-1716536, AST-1312651) and the Gordon and Betty Moore Foundation (GBMF-5278) for financial support of this work. This work was supported in part by the Black Hole Initiative at Harvard University, which is supported by a grant from the John Templeton Foundation.
\end{acknowledgements}

\bibliographystyle{aa}
\bibliography{rgauss.bib}

\begin{appendix}
\section{Properties of the visibility function}\label{sec:vis_prop}

\subsection{Visibility derivatives and image moments}\label{sec:vis_der}

Non-astrometric VLBI experiments such as the EHT measure visibility amplitudes directly but do not provide absolute phase information. Nevertheless, the zeroth and second image moments are determined from visibility amplitudes alone \citep[i.e., they do not depend on the measured phase;][]{moffet_1962,burn_1976}. For instance, the total flux density $\int I(\mathbf{x}) d^2\mathbf{x} = V(\mathbf{0}) = |V(\mathbf{0})|$ because the zero-baseline visibility is real and positive, and therefore equal to its modulus.

More generally, we can express the visibility function as a Taylor expansion of its derivatives:
\begin{align}
V(\mathbf{u}) &= \int d^2\mathbf{x}\ I(\mathbf{x}) \left[ 1 - 2i\pi\mathbf{u}\cdot \mathbf{x} - \frac{(2\pi\mathbf{u}\cdot \mathbf{x})^2}{2} \right. \nonumber  \\
& \qquad \left. + \frac{i(2\pi\mathbf{u}\cdot \mathbf{x})^3}{6} 
+ \frac{(2\pi\mathbf{u}\cdot \mathbf{x})^4}{24} + \cdots \right].
\end{align}
The visibility amplitude function is image-translation invariant. To obtain a Taylor expansion for visibility amplitudes, we choose the image centroid to be at the origin. The first derivative of the visibility function (thus the second term of the Taylor expansion) then vanishes, giving 
\begin{align}
\nonumber V(\mathbf{u}) &\simeq \int d^2\mathbf{x}\ I(\mathbf{x}) \left[ 1  - \frac{(2\pi\mathbf{u}\cdot \mathbf{x})^2}{2}  \right] \\
 &\simeq V(\mathbf{0}) - 2\pi^2 \int d^2\mathbf{x}\ (\mathbf{u}\cdot \mathbf{x})^2 I(\mathbf{x}).
\end{align}
On short baselines (i.e., those with $\mathbf{u}\cdot \mathbf{x} \ll 1$), the visibility function is then positive and real, so $\left| V(\mathbf{u}) \right| \simeq V(\mathbf{u})$.
Since $\mathbf{u} =\begin{pmatrix} u \\ v  \end{pmatrix} $ and $\mathbf{x} =\begin{pmatrix} x \\ y  \end{pmatrix} $, we can expand the inner product of the two vectors:
\begin{align}
\nonumber (\mathbf{u}\cdot \mathbf{x})^2 &= u^2x^2 + v^2y^2 + 2uvxy  \\
    &= \begin{pmatrix} u & v  \end{pmatrix} \begin{pmatrix} x^2 & xy \\ xy & y^2  \end{pmatrix} \begin{pmatrix} u \\ v  \end{pmatrix}.
\end{align}
Combining these results with the definition of the covariance matrix $\boldsymbol{\Sigma}$ (see Appendix~\ref{sec:vis_axes}), we obtain:
\begin{align}
\nonumber |V(\mathbf{u})| &\simeq V(\mathbf{0}) - 2\pi^2  \int d^2\mathbf{x}\, (\mathbf{u}\cdot \mathbf{x})^2 I(\mathbf{x}) \\
\nonumber &\simeq V(\mathbf{0}) - 2\pi^2 \begin{pmatrix} u & v  \end{pmatrix} \int d^2\mathbf{x}\,I(\mathbf{x})\begin{pmatrix} x^2 & xy \\ xy & y^2  \end{pmatrix} \begin{pmatrix} u \\ v  \end{pmatrix} \\
    &\simeq V(\mathbf{0}) - 2\pi^2 V(\mathbf{0})\mathbf{u}^\intercal \boldsymbol{\Sigma} \mathbf{u} . \label{eq:ap_gen-short}
\end{align}
The downward curvature of the amplitude function at zero baseline is thus related to the image covariance by:
\begin{align}
    \left. \nabla\nabla ^\intercal |V(\mathbf{u})| \right\rfloor_{\mathbf{u}=\mathbf{0}} = \left. \nabla\nabla^\intercal V(\mathbf{u}) \right\rfloor_{\mathbf{u}=\mathbf{0}} = -4\pi^2 V(\mathbf{0}) \boldsymbol{\Sigma}. \label{eq:ap_2mom}
\end{align}

\subsection{Image principal axes and visibility curvature}\label{sec:vis_axes}

From Equation~\ref{eq:ap_2mom}, the curvature of the visibility function on short baselines is proportional to the second central moment of the image projected along the baseline direction. The second central moment of the image is naturally expressed as a covariance matrix:
\begin{align}
\mathbf{\Sigma} &\equiv \frac{\int d^2\mathbf{x} I(\mathbf{x}) (\mathbf{x}-\boldsymbol{\mu})(\mathbf{x}-\boldsymbol{\mu})^\intercal}{\int  d^2\mathbf{x}\, I(\mathbf{x}) } = 
\begin{pmatrix} 
\sigxx & \sigxy \\ 
\sigyx & \sigyy 
\end{pmatrix}, \label{eq:ap_matrix} \\ 
\nonumber \sigxx &= \frac{\int  d^2\mathbf{x}\, I(\mathbf{x}) (x-\xo)^2}{\int   d^2\mathbf{x}\, I(\mathbf{x})},\\ 
\nonumber \sigyy &= \frac{\int  d^2\mathbf{x}\, I(\mathbf{x}) (y-\yo)^2}{\int   d^2\mathbf{x}\, I(\mathbf{x})},\\ 
\nonumber \sigxy &= \frac{\int  d^2\mathbf{x}\, I(\mathbf{x}) (x-\xo)(y-\yo)}{\int   d^2\mathbf{x}\, I(\mathbf{x})} = \sigyx.
\end{align}
To put the covariance matrix in a more intuitive form, we express it in terms of its principal axes. The image covariance matrix has two eigenvalues, and can be diagonalized as follows:
\begin{align}
\mathbf{\Sigma} 
= \mathbf{R_\phi} \begin{pmatrix} 
\lmin & 0 \\ 
0 & \lmaj
\end{pmatrix} \mathbf{R_\phi^{\intercal}},  
\end{align}
where the rotation matrix  $\mathbf{R_\phi}$, based on the position angle $\phi$ (East of North) of the major principal axis, is given by:
\begin{align}
\mathbf{R_\phi} = \begin{pmatrix} 
\cos(\phi) & \sin(\phi) \\ 
-\sin(\phi) & \cos(\phi)
\end{pmatrix} .  
\end{align}
The eigenvalues are derived from the quadratic equation:
\begin{align}
\lmaj &= \frac{\sigxx+\sigyy}{2} + \frac{\sqrt{4(\sigxy)^2 + (\sigxx - \sigyy)^2}}{2}, \\
\lmin &= \frac{\sigxx+\sigyy}{2} - \frac{\sqrt{4(\sigxy)^2 + (\sigxx - \sigyy)^2}}{2}. 
\end{align}
We can also express each term of the covariance matrix in terms of the eigenvalues and position angle $\phi$:
\begin{align}
\sigxx &= \cos^2(\phi)\lmin + \sin^2(\phi) \lmaj, \\
\sigyy &= \sin^2(\phi) \lmin + \cos^2(\phi) \lmaj, \\
\sigxy &= (\lmaj - \lmin)\cos(\phi) \sin(\phi).
\end{align}
The eigenvalues of the covariance matrix are the variances along the principal axes (major and minor axes).

\section{Implementation via gradient descent}\label{sec:implementation}
\subsection{Pixel-based derivation of principal axes}

Here we present the computation of the covariance matrix for the pixel-based reconstructions from RML. The centroid of an $n \times n$ pixel-based image is given by the following parameters:
\begin{align}
 {\xo} = \frac{\sumk \xk \Ik}{\sumk \Ik} \mbox{ and }
 {\yo} = \frac{\sumk \yk \Ik}{\sumk \Ik},
\end{align}
where i is the pixel number (from 1 to k), \Ik~is the intensity at that pixel, \xk~is the x-position and \yk~is the y-position of the pixel in the image. The second moment of the image is given by the covariance matrix
 \begin{align}
\mathbf{\Sigma} &=
\begin{pmatrix} 
\sigxx & \sigxy \\ 
\sigxy & \sigyy 
\end{pmatrix} , \\
\mbox{ where }\\
\sigxx &= \frac{\sumk (\xk - \xo)^2 \Ik}{\sumk \Ik},\\
\sigyy &= \frac{\sumk (\yk - \yo)^2 \Ik}{\sumk \Ik}, \\
\sigxy &= \frac{\sumk (\xk - \xo)(\yk-\yo) \Ik}{\sumk \Ik}.
\end{align}
As in Appendix~\ref{sec:vis_axes}, the image covariance matrix has two eigenvalues and can be diagonalized to obtain the principal axes FWHMs.

\subsection{Gradient Descent Implementation}
We have defined our regularization function via the Frobenius norm:
\begin{align}
\mathcal{R}_\mathbf{\Sigma} = (\sigxx - \sigxx')^2 + (\sigyy - \sigyy')^2 + 2(\sigxy - \sigxy')^2 .
\end{align}
Within the framework of the \texttt{eht-imaging} library, the objective function is minimized via gradient descent. Therefore, the regularization functions must also individually be minimized via gradient descent.
The gradients for the quantities describing the properties of the image introduced thus far, for a given pixel j, are given below:
\begin{align}
\nonumber \frac{\delta \xo}{\delta \Ij} &= \frac{\xj \sumk \Ik - \sumk (\xk \Ik)}{\big(\sumk \Ik\big)^2} = \frac{\xj  - \xo}{\big(\sumk \Ik\big)},\\
\frac{\delta \yo}{\delta \Ij} &= \frac{\yj \sumk \Ik - \sumk (\yk \Ik)}{\big(\sumk \Ik\big)^2} = \frac{\yj  - \yo}{\big(\sumk \Ik\big)},
\end{align}

\begin{align}
 \nonumber \frac{\delta \sigxx}{\delta \Ij} &= 
\frac{[(\xj - \xo)^2 - 2(\xj - \xo)\frac{\delta \xo}{\delta \Ij} \Ij] \sumk \Ik - \sumk [(\xk - \xo)^2 \Ik]}{\big(\sumk \Ik\big)^2} \\
&= \frac{[(\xj - \xo)^2 - 2(\xj - \xo)\frac{\delta \xo}{\delta \Ij} \Ij]- \sigxx}{\sumk \Ik},
\end{align}

\begin{align}
\nonumber \frac{\delta \sigyy}{\delta \Ij} &=
\frac{[(\yj - \yo)^2 - 2(\yj - \yo)\frac{\delta \yo}{\delta \Ij} \Ij] \sumk \Ik - \sumk [(\yk - \yo)^2 \Ik]}{\big(\sumk \Ik\big)^2} \\
&= \frac{[(\yj - \yo)^2 - 2(\yj - \yo)\frac{\delta \yo}{\delta \Ij} \Ij] - \sigyy}{\sumk \Ik},
\end{align}

\begin{align}
\nonumber &\frac{\delta \sigxy}{\delta \Ij} = \frac{[(\xj - \xo)(\yj - \yo) - (\yj - \yo)\frac{\delta \xo}{\delta \Ij} \Ij] \sumk \Ik }{\big(\sumk \Ik\big)^{2}} \\
\nonumber &- \frac{[ (\xj - \xo)\frac{\delta \yo}{\delta \Ij} \Ij] \sumk \Ik }{\big(\sumk \Ik\big)^{2}}- \frac{ \sumk [(\xk - \xo)(\yk - \yo) \Ik]}{\big(\sumk \Ik\big)^{2}} \\
&= \frac{[(\xj - \xo)(\yj - \yo) - (\yj - \yo)\frac{\delta \xo}{\delta \Ij} \Ij - (\xj - \xo)\frac{\delta \yo}{\delta \Ij} \Ij]- \sigxy}{\sumk \Ik} 
\end{align}

We can now compute the gradient of the second moment regularization within the minimization framework of the \texttt{eht-imaging} library:
\begin{align}
\nonumber \frac{\delta \mathcal{R}_\Sigma}{\delta \Ij} = 2(\sigxx - \sigxx')\frac{\delta \sigxx}{\delta \Ij} + 2(\sigyy - \sigyy')\frac{\delta \sigyy}{\delta \Ij} \\
+ 4(\sigxy - \sigxy')\frac{\delta \sigxy}{\delta \Ij} .
\end{align}

Note that these equations correspond to regularization of the normalized second central moment of an image. In cases where the total flux density of an image is constrained or regularized, it would be advantageous to instead regularize the unnormalized second central moment, giving a substantially simplified and convex optimization problem.

\end{appendix}
\end{document}